# A Panchromatic JWST Spectrum of a Giant Starspot on the Fully Convective M Dwarf TOI-3884

C. A. Murray[1], L. Garcia[2], B. V. Rackham[3,4], Z. Berta-Thompson[1], A. D. Feinstein[5], S. J. Mercier[3], B. Charnay[6,7], L. Hebb[8], J. E. Libby-Roberts[9,10,11], Y. Rotman[12], A. Stephens[1], M. Timmermans[13,14], L. Welbanks[12], K. Barkaoui[3,14,15], Caleb I. Cañas[16,17], M. Delamer[9,10], E. Ducrot[6,18], S. Kanodia[19], S. Mahadevan[9,10], J. P. Ninan[20], and J. de Wit[3]

[1] Department of Astrophysical and Planetary Sciences, University of Colorado Boulder, Boulder, CO 80309, USA; Catriona.Murray@colorado.edu
[2] Center for Computational Astrophysics, Flatiron Institute, New York, NY, USA
[3] Department of Earth, Atmospheric and Planetary Sciences, Massachusetts Institute of Technology, 77 Massachusetts Avenue, Cambridge, MA 02139, USA
[4] Kavli Institute for Astrophysics and Space Research, Massachusetts Institute of Technology, Cambridge, MA 02139, USA
[5] Department of Physics and Astronomy, Michigan State University, East Lansing, MI 48824, USA
[6] LIRA, Observatoire de Paris, Université PSL, CNRS, Sorbonne Université, Université Paris Cité, 5 place Jules Janssen, Meudon, 92195, France
[7] Laboratoire d'Astrophysique de Bordeaux, Univ. Bordeaux, CNRS, B18N, allée Geoffroy Saint-Hilaire, Pessac, 33615, France
[8] Hobart and William Smith Colleges, Geneva, NY 14456, USA
[9] Department of Astronomy & Astrophysics, 525 Davey Laboratory, The Pennsylvania State University, University Park, PA 16802, USA
[10] Center for Exoplanets and Habitable Worlds, The Pennsylvania State University, University Park, PA 16802, USA
[11] Department of Physics and Astronomy, University of Tampa, Tampa, FL 34606, USA
[12] School of Earth and Space Exploration, Arizona State University, Tempe, AZ 85281, USA
[13] School of Physics & Astronomy, University of Birmingham, Edgbaston, Birmingham B15 2TT, UK
[14] Astrobiology Research Unit, Université de Liège, 19C Allée du 6 Août, 4000 Liège, Belgium
[15] Instituto de Astrofísica de Canarias (IAC), Calle Vía Láctea s/n, 38200, La Laguna, Tenerife, Spain
[16] Southeastern Universities Research Association, Washington, DC 20005, USA
[17] NASA Goddard Space Flight Center, 8800 Greenbelt Road, Greenbelt, MD 20771, USA
[18] Université Paris-Saclay, Université Paris Cité, CEA, CNRS, AIM, Gif-sur-Yvette, France
[19] Earth and Planets Laboratory, Carnegie Institution for Science, 5241 Broad Branch Road, NW, Washington, DC 20015, USA
[20] Department of Astronomy and Astrophysics, Tata Institute of Fundamental Research, Homi Bhabha Road, Colaba, Mumbai 400005, India

## Abstract

TOI-3884 b is a rare super-Neptune transiting a fully convective M dwarf that hosts a persistent giant polar spot. Because the planet occults this active region during every transit, the system offers a unique laboratory to directly probe the stellar surface and spot properties. We present seven James Webb Space Telescope (JWST) transits of TOI-3884 b observed with NIRISS and NIRSpec (spanning 0.6–5.3 $\mu$m. While all visits show a recurring spot-crossing signature, each transit exhibits a distinct spot-crossing morphology, enabling us to infer a stellar rotation period of $P = 11.102 \pm 0.003$ days and tightly constrain the pole-on stellar orientation ($i_* = 139\overset{\circ}{.}2 \pm 0\overset{\circ}{.}3$, $\lambda_* = 31\overset{\circ}{.}1 \pm 0\overset{\circ}{.}4$, $\psi_* = 56\overset{\circ}{.}4 \pm 0\overset{\circ}{.}3$) and spot properties ($R_{\rm spot} = 0.576\ ^{+0.006}_{-0.005}\ R_*$, $\phi_{\rm spot} = -84\overset{\circ}{.}69 \pm 0\overset{\circ}{.}12$). We leverage this orbital configuration to measure the first empirical panchromatic spectrum of an M dwarf starspot with JWST, establishing a direct observational benchmark for stellar atmosphere models in the fully convective regime. Comparison with 1D NewEra and SPHINX atmosphere models indicates that the spot is $183 \pm 1$ K cooler than the photosphere, consistent with previous ground-based measurements and expectations for mid M dwarf spot contrasts. While the models reproduce the observed contrasts at wavelengths longer than 1 $\mu$m, they significantly underpredict the contrasts at shorter wavelengths. These results demonstrate that M dwarf stellar atmosphere models may not fully capture the wavelength dependence of stellar contamination in transmission spectra and highlight the importance of empirical spot spectra for robust interpretation of planetary atmospheres, particularly in the optical.

## 1. Introduction

One of the most urgent challenges of accurately interpreting planetary transmission spectra observations around M dwarfs is dealing with stellar surface heterogeneity. Stars are speckled with cooler (spots) and hotter (faculae, plages) regions that imprint wavelength-dependent signals on transit depths, a phenomenon known as the "transit light source (TLS) effect" (B. V. Rackham et al. 2018, 2019). For low-mass stars with enhanced activity and large spot-covering fractions, the TLS effect can mask or mimic molecular features in transmission spectra (e.g., B. Rackham et al. 2017; N. Espinoza et al. 2019; O. Lim et al. 2023; E. M. May et al. 2023). Even relatively quiet M dwarfs can exhibit highly heterogeneous surfaces, as shown by photometric monitoring of slowly rotating M dwarfs (E. R. Newton et al. 2016, 2017). Furthermore, recent Hubble

Space Telescope (HST) and James Webb Space Telescope (JWST; J. P. Gardner et al. 2006) studies of small exoplanets around cool stars (e.g., GJ 486 b; S. E. Moran et al. 2023, GJ 1132b; E. M. May et al. 2023; TOI 270 d T. Mikal-Evans et al. 2023, L 98-59 c; T. Barclay et al. 2025, LHS 1140b; C. Cadieux et al. 2024, and the TRAPPIST-1 planets; J. de Wit et al. 2016, 2018; Z. Zhang et al. 2018; O. Lim et al. 2023; N. Espinoza et al. 2025; A. Glidden et al. 2025; C. Piaulet-Ghorayeb et al. 2025; M. Radica et al. 2025; A. D. Rathcke et al. 2025; N. H. Allen et al. 2026) have struggled to definitively separate absorption from the planet's atmosphere and from cool stellar spots.

Current mitigation strategies typically handle stellar contamination by fitting for TLS signals as a function of the temperature contrast, $C$, between active regions and the quiescent photosphere, and the surface covering fraction of those regions, $f$. In practice, $C$ is approximated using stellar atmosphere models at different temperatures to represent the photosphere and active regions. However, current 1D non-magnetic stellar models for M dwarfs (such as PHOENIX; T. O. Husser et al. 2013, SPHINX; A. R. Iyer et al. 2023, NewEra; P. H. Hauschildt et al. 2025) exhibit discrepancies that can lead to inconsistent atmospheric inferences when correcting for TLS effects on transmission spectra. For example, A. R. Iyer & M. R. Line (2020) showed that differences between generations of PHOENIX models significantly biased derived planetary irradiation temperatures, metallicities, and carbon-to-oxygen ratios, while B. V. Rackham & J. de Wit (2024) found that discrepancies between PHOENIX and SPHINX models dominated the noise budget, inflating uncertainties by up to 50%. Additionally, these models often are a poor match to empirical spectra (H. R. Wakeford et al. 2019; L. J. Garcia et al. 2022; O. Lim et al. 2023; F. Davoudi et al. 2024). Recent 3D magnetohydrodynamic (MHD) simulations of M dwarfs offer some hope of reconciling some of these disparities (H. N. Smitha et al. 2025), but these predictions remain largely untested against empirical panchromatic measurements. The absence of direct starspot spectra for M dwarfs is therefore a critical gap in interpreting transmission spectra in the JWST era.

Spot-crossing events provide a direct observational pathway to address this problem (B. V. Rackham et al. 2023). When a planet transits over an active feature, the resulting light-curve signature can constrain the spot's size, location, and spectral contrast (e.g., K. F. Huber et al. 2010; D. K. Sing et al. 2011; F. Pont et al. 2013; L. Mancini et al. 2017; B. M. Morris et al. 2017; N. Espinoza et al. 2019). If the spots share a common temperature, these events in turn reveal information about the unocculted stellar surface, enabling an empirical estimate of TLS signals as a necessary benchmark for theoretical models and as a physically grounded foundation for TLS corrections.

TOI-3884 b offers a uniquely powerful laboratory for probing M dwarf spot spectra. The planet is a short-period ($P = 4.54$ days) super-Neptune ($6.4 \pm 0.2\,R_{\oplus}$, $32.6 \pm 7.3\,M_{\oplus}$) transiting a nearby ($d = 43$ pc) fully convective M dwarf ($M = 0.30\,M_{\odot}$, $T_{\rm eff} = 3180 \pm 88$ K; G. Chabrier & I. Baraffe 1997; J. E. Libby-Roberts et al. 2023). All observed transits exhibit a large, repeatable spot-crossing anomaly (J. M. Almenara et al. 2022; J. E. Libby-Roberts et al. 2023), while long-term photometry shows minimal, though detectable, 11 days rotational modulation (M. Mori et al. 2025; P. Tamburo et al. 2025). Together, these observations indicate a nearly pole-on stellar inclination and a long-lived, near-polar active region that is occulted during every transit (J. E. Libby-Roberts et al. 2023; H. Chakraborty et al. 2025; M. Mori et al. 2025; S. Sagynbayeva et al. 2025; P. Tamburo et al. 2025). This long lifetime may reflect the reduced differential rotation expected in fully convective M dwarfs, for which lower surface temperatures and long convective-turnover timescales inhibit the shear that typically disrupts starspots (M. Küker & G. Rüdiger 2008; L. L. Kitchatinov & S. V. Olemskoy 2011).

Spectropolarimetric surveys (J.-F. Donati et al. 2006; C. Moutou et al. 2007; J. Morin et al. 2008, 2010) have shown that fast-rotating M dwarfs (with periods comparable to TOI-3884) can host strong, stable magnetic fields with large poloidal components, consistent with giant polar spots. Though these "polar spot caps" may be common, TOI-3884's polar geometry is exceptionally rare for transiting exoplanets: it virtually guarantees spot occultations and allows the planet to repeatedly probe the same active region. The result is a natural laboratory for empirically measuring the spectral properties of a starspot on a fully convective M dwarf.

Giant planets are intrinsically rare around low-mass M dwarf stars (G. Laughlin et al. 2004; S. Ida & D. N. C. Lin 2005; J. C. Morales et al. 2019; R. Burn et al. 2021)—TOI-3884 b is the only known super-Neptune planet to orbit a fully convective M dwarf (cooler than 3300 K or smaller than $0.35\,M_{\odot}$; G. Chabrier & I. Baraffe 1997). Their deep transits and strong atmospheric features make these planets sensitive probes of ongoing atmospheric processes (e.g., aerosol formation, disequilibrium chemistry, or atmospheric escape), and characterizing their atmospheres accurately can help us understand an extreme limit of gas giant formation and evolution.

In this Letter, we present seven JWST transit observations of TOI-3884 b spanning 0.6–5.3 $\mu$m and use the recurrent spot-crossing anomalies to derive the first empirical panchromatic spectrum of a starspot on a fully convective M dwarf. Section 2 describes the observations and data reductions. In Section 3, we derive wavelength-independent stellar and orbital parameters from the NIRSpec data (Section 3.1), followed by a self-consistent joint model of the six visits least impacted by flares to constrain the stellar orientation and rotation and spot geometry (Section 3.2). We cross-check our results by performing individual visit fits in Section 3.3. We present the resulting spot contrast spectrum and compare it to stellar models in Section 4, place our results in a broader context in Section 5, and summarize our conclusions in Section 6.

## 2. Observations and Data Reduction

### 2.1. JWST Observations

We combine seven transit observations of TOI-3884 b from JWST Cycle 3 programs GO-5863 (PI: Murray) and GO-5799 (PI: Garcia, co-PIs: Rackham, Timmermans, Charnay),[21] summarized in Table 1. Due to the similarity in science goals and observing strategy between the programs, we opted to combine teams and observations to produce a more powerful dataset.

#### 2.1.1. GO-5863

In the GO-5863 program, we observed four consecutive transits of TOI-3884 b, alternating between NIRISS (L. Albert et al. 2023; R. Doyon et al. 2023) in Single Object Slitless

[21] All the JWST data used in this Letter can be found on MAST doi:10.17909/bn1d-sz91.

**Table 1**
Dates and Instrument Modes for the Seven JWST Visits from the GO-5863 and GO-5799 Programs

| Visit | Date | Time (days) | Instrument | $N$ [a] |
|---|---|---|---|---|
| 1 | 2024 Dec 16 | 60660.2 | NIRISS/SOSS | 1 |
| 2 | 2024 Dec 20 | 60664.7 | NIRSpec/G395M | 2 |
| 3 | 2024 Dec 25 | 60669.3 | NIRISS/SOSS | 3 |
| 4 | 2024 Dec 29 | 60673.9 | NIRSpec/G395M | 4 |
| 5 | 2025 May 24 | 60819.3 | NIRSpec/G395H | 36 |
| 6 | 2025 Jun 2 | 60828.3 | NIRISS/SOSS | 38 |
| 7 | 2025 Jun 15 | 60842.0 | NIRISS/SOSS | 41 |

**Note.** Throughout this work, times are defined in BJD (in TDB)-2400000.5.
[a] Transit index $N$, defined such that $N = 1$ is the first JWST observation, incremented by one for each orbit of TOI-3884 b.

Spectroscopy (SOSS) mode (Visit 1: 2024 December 16, Visit 3: 2024 December 25) and NIRSpec (S. M. Birkmann et al. 2022; P. Jakobsen et al. 2022) in Bright Object Time (BOTS) mode with the G395M grating (Visit 2: 2024 December 20, Visit 4: 2024 December 29).

For NIRISS/SOSS, we utilized the GR700XD disperser with the SUB256 subarray and NISRAPID readout pattern. We used 10 groups/integration (60.4 s/integration, 96 integrations in transit) to remain below ∼80% of saturation while maintaining a cadence sufficient to resolve short-timescale stellar variability. Each visit consisted of 251 science integrations over ∼4.2 hr, corresponding to the transit duration (1.60 hr) with an equal amount of out-of-transit baseline plus an additional 1 hr buffer for scheduling flexibility and instrument settling. Target acquisition was performed with SOSSFAINT and 19 groups.

For NIRSpec, we used BOTS mode with the G395M grating, F290LP filter, S1600A1 slit ($1\farcs6 \times 1\farcs6$), SUB2048 subarray, and NRSRAPID readout pattern. We observed 23 groups/integration, giving 21.6 s/integration and 267 integrations in transit. For target acquisition, we relied on JWST's natural pointing ($\sim 0\farcs1$ radial, well within the aperture of the $1\farcs6 \times 1\farcs6$ S1600A1 slit) due to the lack of suitable unsaturated reference sources within the visit splitting distance.

#### 2.1.2. GO-5799

In the GO-5799 program, we observed three transits of TOI-3884 b: one with NIRSpec/G395H (Visit 5: 2025 May 24), followed by two with NIRISS/SOSS (Visit 6: 2025 June 2, Visit 7: 2025 June 15).

The NIRISS/SOSS observations used the same setup as in GO-5863, with 10 groups/integration. Each visit consisted of 321 integrations, yielding a total exposure time of 5.4 hr per visit. Target acquisition was performed on TOI-3884 using the SOSSFAINT mode with the F480M filter, NISRAPID readout pattern, and three groups.

The NIRSpec observation followed the same BOTS mode setup as GO-5863 but with the G395H grating. To remain below ∼80% of saturation, we used 70 groups per integration, resulting in 303 integrations and a total exposure time of 5.4 hr. Target acquisition was performed using the Wide Aperture Target Acquisition mode.

### 2.2. Data Reduction

All visits were reduced independently using two pipelines: `Eureka!` (v1.2, T. J. Bell et al. 2022)[22] and the EXOplanet Transit and Eclipse Data Reduction Framework (`ExoTEDRF`; formerly `supreme-SPOON`, A. D. Feinstein et al. 2023; M. Radica 2024; M. Radica et al. 2023)[23]. The two reductions yielded consistent spectroscopic light curves within uncertainties; unless otherwise noted, we adopt the `ExoTEDRF` reductions for subsequent analysis. We briefly summarize these reductions here and provide full reduction details in Appendix A.

`Eureka!` performs detector-level corrections using the JWST calibration pipeline, followed by time series-optimized extraction. We applied standard Stage 1 and 2 calibrations, including ramp fitting, dark subtraction, flat-fielding, and wavelength calibration, with adjustments appropriate for time series data (e.g., group-level background correction and tuned jump rejection thresholds). Spectra were extracted using optimal extraction after trace identification and background subtraction. NIRSpec/G395H NRS1 and NRS2 data were reduced separately.

`ExoTEDRF` provides an independent end-to-end reduction framework for JWST exoplanet time series. We performed standard detector calibrations using the JWST pipeline, including superbias subtraction, dark correction, flat-fielding, and wavelength calibration. For NIRISS/SOSS, we scaled and subtracted the STScI background model and applied a group-level $1/f$ noise correction. For NIRSpec, additional integration-level $1/f$ correction and wavelength adjustments were applied as needed. Spectra were extracted using box apertures optimized for each instrument mode.

### 2.3. Qualitative Behavior of the Reduced Light Curves

Figures 1 and 2 present the reduced JWST transit light curves of TOI-3884 b, shown respectively as spectroscopic and white-light time series for all seven visits. In all visits, a prominent spot-crossing event is apparent during the first half of the transit, suggesting the polar spot persists over the full observational baseline. The timing, width, and detailed morphology of the feature vary between visits, suggesting changes in the projected location of the occulted region due to stellar rotation and possible substructure or intrinsic evolution within the active region.

Several visits also show clear signatures of stellar flaring. Visits 1–5 exhibit flares before, during, or after transit (Figure 2), with the first two NIRISS observations (Visits 1 and 3) particularly affected by strong pre-ingress outbursts of multiple flares that extend into transit. Visit 3 has an unusual flare shape, characterized by a sharp rise followed by an approximately linear decline rather than the usual exponential decay. This may reflect overlapping flare events, including one that occurs during ingress (around $t = 60669.42$ days in Figure 6 in Appendix B). The NIRISS bandpass covers a number of chromospheric activity indicators—such as H$\alpha$ (0.656 $\mu$m), several Paschen (H8-3, 0.9546 $\mu$m, H7-3, 1.005 $\mu$m, H6-3, 1.094 $\mu$m, H5-3, 1.282 $\mu$m, H4-3, 1.875 $\mu$m) and Brackett (H8-4, 1.944 $\mu$m) lines, and the He I and Ca II infrared triplets (1.0833 $\mu$m and 0.85–0.87 $\mu$m respectively)—which show enhanced emission during flares, confirming their stellar nature. The H$\alpha$ light curves for Visits 1 and 3 are shown in Figure 6 in Appendix B.

[22] https://eurekadocs.readthedocs.io/en/latest/
[23] https://exotedrf.readthedocs.io/en/latest/

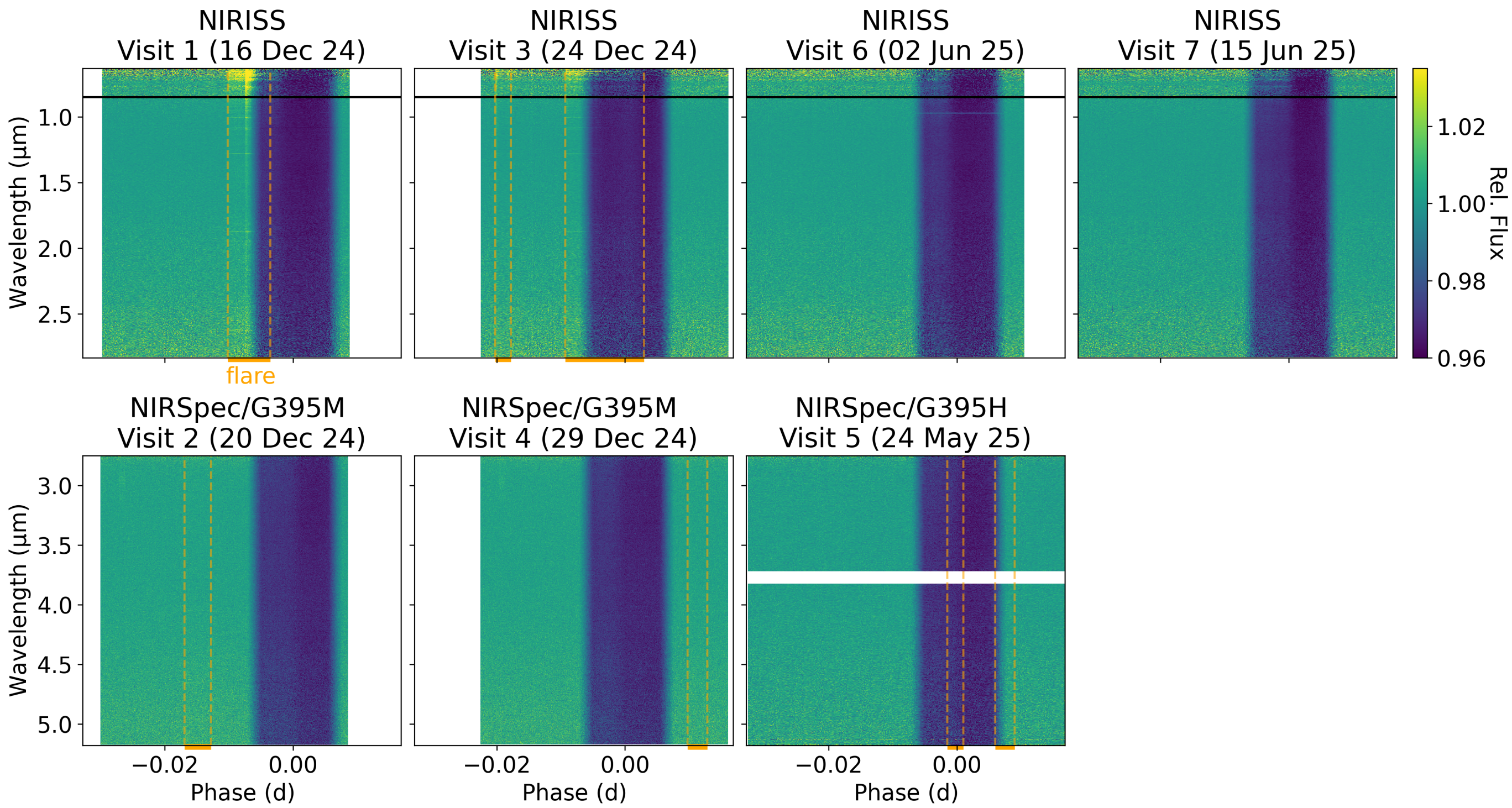

**Figure 1.** Spectroscopic time series of the transit light curves of TOI-3884 b for all seven JWST visits, reduced with ExoTEDRF (Section 2.2). NIRISS/SOSS observations are presented in the top row, and NIRSpec/BOTS in the bottom. Each panel shows normalized flux as a function of orbital phase and wavelength, and the color scale is shared across all panels. In all visits, the planetary transit appears as a broad vertical band centered at phase zero, and the spot-crossing event is apparent in the first half of the transit. Several visits also exhibit continuum and line emission attributable to stellar flares. The horizontal black lines (0.85 $\mu$m) on the top panels indicate approximately where NIRISS orders 1 and 2 overlap (0.85–1.10 $\mu$m). The white horizontal bar in Visit 5 indicates the gap between the NRS1 and NRS2 detectors for the G395H mode (3.72–3.82 $\mu$m); in the G395M mode, only NRS1 is illuminated. The timings of identified flares are marked in orange on the $x$-axes.

Visit 6 (2025 June 2) has a zeroth-order contaminant overlapping order 1 at $\lambda \approx 0.96$–$0.985\,\mu$m in the raw images. This contamination sharply reduces the transit depth at these wavelengths (Figure 1). We mask this region in our spectroscopic light curves.

Aside from a lower first integration in several visits, the white light curves show no significant ramps or long-term slopes (Figure 2). We therefore do not include any polynomial systematics models in our transit fits.

## 3. Light-curve Fitting

Due to the complexity of these light curves and the difficulty in simultaneously fitting large numbers of covariant parameters with Markov Chain Monte Carlo (MCMC) sampling, we fit our observations iteratively, as described in the following sections:

3.1 We first fit only NIRSpec light curves to extract wavelength-independent orbital parameters.
3.2 Keeping orbital geometry fixed, we then jointly fit six light curves with a self-consistent spot model to extract spot properties, stellar geometry, and rotation.
3.3 Finally, we follow a method similar to that in Section 3.2, but fit the spot in each light curve independently to check for consistency.

We opted to derive the orbital parameters from NIRSpec observations alone in Section 3.1 because of potential stellar activity biases in NIRISS. The effect of limb darkening and the contrasts between flares, spots, and the stellar photosphere all decrease toward redder wavelengths, meaning that NIRSpec observations ($\lambda = 3.0$–$5.3\,\mu$m) are significantly less impacted by the star than for NIRISS, thus yielding more accurate orbital parameters. Repeating our analysis in Section 3.3 for each visit separately, where we do not enforce a single static spot model, allows for a more flexible spot model that can capture imperfections in our spot geometry as well as true evolution.

### 3.1. Extracting Orbital Parameters

To determine wavelength-independent stellar and orbital parameters, we fit only the three JWST/NIRSpec visits (2, 4, and 5).

#### 3.1.1. Combining the NIRSpec Visits

Visits 2 and 4 used the medium-resolution G395M grating, while Visit 5 was observed with the higher-resolution G395H grating. To combine all NIRSpec visits consistently, we extracted only the G395H NRS1 wavelength range (2.75–3.72 $\mu$m) for our first two visits, essentially creating mock NRS1 light curves. We then binned, normalized, and stitched all three visits together consecutively into a single white light curve with three transits for joint modeling.

#### 3.1.2. Treatment of NIRSpec Flares

All three NIRSpec observations exhibit minor flare events. Visit 2 shows one pre-transit flare, Visit 4 one post-transit, and Visit 5 two small flares: one during the spot-crossing and one post-egress. The flares in Visit 5 are independently confirmed in simultaneous Gemini/GMOS observations (Program GN-2025A-DD-108, PI: Murray; A. Stephens et al. 2026, in preparation).

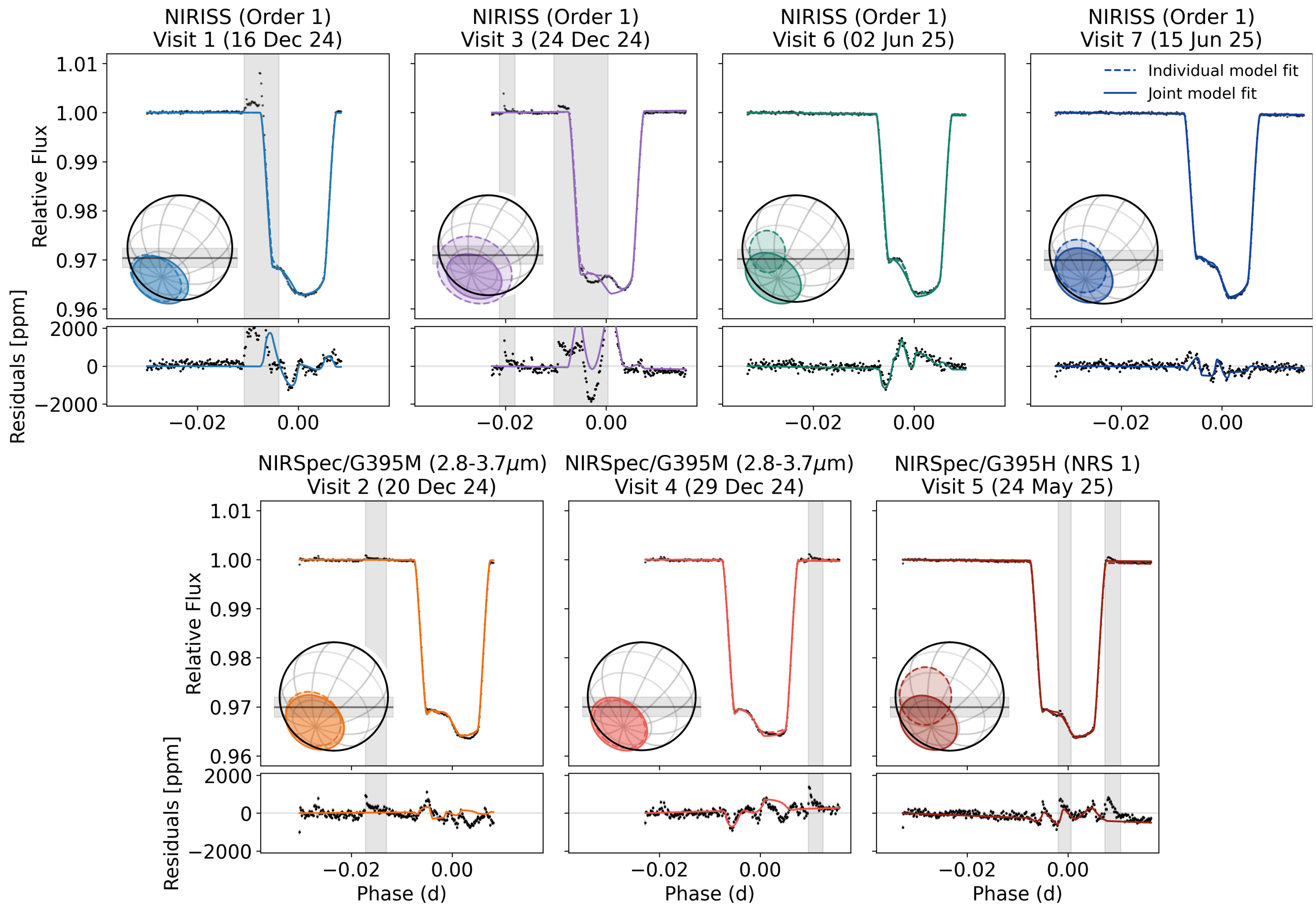

**Figure 2.** White-light transit light curves of TOI-3884 b for all seven JWST visits using the ExoTEDRF reduction described in Section 2.2. NIRISS/SOSS observations are presented in the top row, and NIRSpec/BOTS ($\lambda = 2.8$–$3.7$ $\mu$m to match G395H NRS1) in the bottom. For each visit, the upper panel shows the normalized flux as a function of transit phase as well as the best-fitting white-light models for each visit independently (Section 3.3, dashed lines), and for the joint fit excluding Visit 3 (Section 3.2, solid lines). We plot how the respective individual and joint spot fits for each visit would look on TOI-3884 in each panel. The lower panels show the corresponding flux residuals for the joint model with black points (data point uncertainties are plotted but are almost negligible compared to the residual structure), and the difference between the individual and joint models with the colored lines. For some visits, modeling the light curve individually reduces residuals (e.g., Visit 6); however, most visits show significant structure regardless of the fitting method used. All visits exhibit a pronounced spot-crossing feature during the first half of the transit, though its shape clearly varies between visits. Stellar flares, marked by shaded gray regions, are evident in most light curves. Visits 1 and 3, in particular, show strong stellar flares that overlap the transit window. The NIRSpec observations, which probe longer wavelengths, show comparatively quiescent behavior, though flares are still present in all NIRSpec datasets.

We masked the out-of-transit flares in Visits 2 and 4 to reduce the number of fitted parameters. For Visit 5, we modeled both small, in-transit flares to save precious in-transit data. We use the G. T. Mendoza et al. (2022) template to model flares as a convolution of a Gaussian with a double exponential. We fit both flares for their epochs $t_{\rm flare}$, amplitudes $A_{\rm flare}$, and full widths at half-maximum $\Delta t_{\rm flare}$, simultaneously with the transit and spot parameters described in the following sections (Table 2).

### 3.1.3. Fitting Spot Crossings

We modeled all spot-crossing events using the Python tool `fleck` (B. Morris 2020), similarly to M. Mori et al. (2025). `fleck` simulates starspots as circular dark regions on the surfaces of rotating stars, accounting for limb darkening, limb foreshortening, and rotational modulation. `fleck` is usually recommended for smaller spots due to known edge effects near the stellar limb,[24] and we discuss the impact of this limitation in Section 5.4. `Fleck`, like most spot-fitting models, also neglects the impact of differential limb darkening across spots.

We note that versions of `fleck` $\leqslant$ 1.1.3 have a sign error in the spin–orbit angle implementation. Therefore, we make the following correction to our angles: $i_* = 180 - i_{\rm fleck}$ and $\lambda_* = 180 - \lambda_{\rm fleck}$ for stellar inclination, $i_*$, and sky-projected obliquity, $\lambda_*$.

### 3.1.4. Parallel-tempered Sampling

Spot properties are inherently degenerate. Geometric symmetries above and below the transit chord and different spot configurations can produce nearly identical light-curve signatures, resulting in a highly multimodal posterior. Standard MCMC methods can get stuck in local maxima in such cases (D. Foreman-Mackey et al. 2013; D. W. Hogg & D. Foreman-Mackey 2018; Section 6.3 of J. Dunkley et al. 2005).

[24] https://fleck.readthedocs.io/en/stable/fleck/details.html

**Table 2**
Derived and Fixed White Light-curve Parameters for the Combined Three JWST/NIRSpec Visits and the Combined NIRSpec and NIRISS Visits

| Parameter | Units | Prior[a] | NIRSpec White Light Curves Only | Joint White Light Curves |
|---|---|---|---|---|
| Stellar mass | $M_*[M_\odot]$ | ⋯ | 0.298[b] | 0.298[b] |
| Stellar rotation period | $P_{\rm rot}$ [days] | $\mathcal{U}(10,\ 12)$ | 11.020[c] | $11.102 \pm 0.003$[d] |
| QLD | $(q_1,\ q_2)_{\rm NIRISS}$ | D. M. Kipping (2013) | ⋯ | $(0.081 \pm 0.005,\ 0.19 \pm 0.03)$[d] |
| QLD | $(q_1,\ q_2)_{\rm NIRSpec}$ | D. M. Kipping (2013) | $(0.029 \pm 0.004,\ 0.017\ ^{+0.026}_{-0.013})$[d] | $(0.030 \pm 0.002,\ 0.04 \pm 0.03)$[d] |
| Stellar inclination | $i_*$ [deg] | $\mathcal{U}(90,\ 180)$ | $154.6 \pm 0.9$[d] | $139.2 \pm 0.3$[d] |
| Sky-projected stellar obliquity | $\lambda_*$ [deg] | $\mathcal{U}(0,\ 360)$ | $37.7 \pm 1.2$[d] | $31.1 \pm 0.4$[d] |
| True obliquity | $\psi_*$ [deg] | ⋯ | $70.6 \pm 0.8$[d] | $56.4 \pm 0.3$[d] |
| Radius ratio | $(R_{\rm p}/R_*)_{\rm NIRISS}$ | $\mathcal{U}(0.1,\ 0.3)$, $\mathcal{N}(0.20,\ 0.05)$ | ⋯ | $0.18237^{+0.00011}_{-0.00012}$[d] |
| Radius ratio | $(R_{\rm p}/R_*)_{\rm NIRSpec}$ | $\mathcal{U}(0.1,\ 0.3)$, $\mathcal{N}(0.20,\ 0.05)$ | $0.18353 \pm 0.00016$[d] | $0.18290 \pm 0.00008$[d] |
| Orbital period | $P_{\rm orb}$ [days] | $\mathcal{N}(4.54,\ 0.10)$ | $4.5445836 \pm 0.0000005$[d] | [e] |
| Semimajor axis | $a/R_*$ | $\mathcal{N}(25,\ 3)$ | $24.73 \pm 0.04$[d] | [e] |
| Transit epoch | $t_0$ [days] | $\mathcal{N}(60664.8947,\ 10 \cdot t_{\rm exp})$ | $60664.894855 \pm 0.000012$[d] | [e] |
| Orbital inclination | $i$ [deg] | $\mathcal{N}(89.7,\ 1.0)$ | $89.54 \pm 0.02$[d] | [e] |
| Eccentricity | $e$ | ⋯ | 0[b] | 0[b] |
| Spot contrast[f] | $C_{\rm NIRISS}$ | $\mathcal{U}(0.0,\ 1.0)$ | ⋯ | $0.189 \pm 0.002$[d] |
| Spot contrast[f] | $C_{\rm NIRSpec}$ | $\mathcal{U}(0.0,\ 1.0)$ | $0.1322 \pm 0.0011$[d] | $0.1396^{+0.0012}_{-0.0010}$[d] |
| Spot radius | $R_{\rm spot}$ $[R_*]$ | $\mathcal{U}(0.0,\ 0.8)$ | $0.480 \pm 0.004$[d] | $0.576\ ^{+0.006}_{-0.005}$[d] |
| Spot latitude | $\phi_{\rm spot}$ [deg] | $\mathcal{U}(-90,\ 0)$ | $-76.6 \pm 0.9$[d] | $-84.69 \pm 0.12$[d] |
| Spot longitude ($t = t_0$) | $\lambda_{\rm spot}$ [deg] | $\mathcal{U}(-180,\ 180)$ | $97.9 \pm 1.3$[d] | $32.60^{+0.17}_{-0.08}$[d] |
| Flare 1 epoch | $t_{\rm flare,1}$ [days] | $\mathcal{N}(60819.407,\ 0.010)$ | $60819.40332^{+0.00013}_{-0.00007}$[d] | [e] |
| Flare 1 amplitude | $A_{\rm flare,1}$ | $\mathcal{U}(0,\ 1)$ | $0.0013\ ^{+0.0002}_{-0.0005}$[d] | [e] |
| Flare 1 FWHM | $\Delta t_{\rm flare,1}$ [days] | $\mathcal{U}(0,\ 1)$ | $0.0012 \pm 0.0004$[d] | [e] |
| Flare 2 epoch | $t_{\rm flare,2}$ [days] | $\mathcal{N}(60819.447,\ 0.010)$ | $60819.4466 \pm 0.0002$ [d] | [e] |
| Flare 2 amplitude | $A_{\rm flare,2}$ | $\mathcal{U}(0,\ 1)$ | $0.00119 \pm 0.00008$[d] | [e] |
| Flare 2 FWHM | $\Delta t_{\rm flare,2}$ [days] | $\mathcal{U}(0,\ 1)$ | $0.0075 \pm 0.0009$[d] | [e] |
| Uncertainty Inflation | $\ln \sigma_{\rm NIRISS}$ | $\mathcal{U}(\ln(0.01),\ \ln(100))$ | ⋯ | $2.51^{+0.03}_{-0.02}$[d] |
| Uncertainty Inflation | $\ln \sigma_{\rm NIRSpec}$ | $\mathcal{U}(\ln(0.01),\ \ln(100))$ | $0.85 \pm 0.03$[d] | $0.67 \pm 0.02$[d] |

**Notes.** For clarity, all times used are in BJD (in TDB)-2400000.5. We recommend the parameters from the "joint white light curves" fit (last column) rather than the "NIRSpec only" as the six observations cover a wider range of stellar rotation phases, placing much stronger constraints on the stellar orientation, rotation, and spot properties.

[a] $\mathcal{U}(A,\ B)$ denotes a uniform prior from $A$ to $B$, and $\mathcal{N}(A,\ B)$ denotes a Gaussian prior with mean $A$ and standard deviation $B$.

[b] J. E. Libby-Roberts et al. ( 2023).

[c] P. Tamburo et al. (2025).

[d] Derived in this work.

[e] Fixed to the value from the NIRSpec-only fit.

[f] In this work, we define contrast as: $C = 0$ when the spot has the same temperature as the quiet photosphere and $C = 1$ when the spot is perfectly dark ($T = 0$ K). We note that `fleck` uses the opposite definition.

To mitigate this, we utilized parallel-tempering MCMC, which runs ensembles of Markov chains at different "temperatures," effectively flattening the likelihood at higher temperatures to enable efficient exploration of parameter space (R. H. Swendsen & J.-S. Wang 1986; C. J. Geyer 1991; D. J. Earl & M. W. Deem 2005; W. D. Vousden et al. 2016). Periodic swaps between chains further improve mixing across modes. Following S. Sagynbayeva et al. (2025), we implemented this approach using `ptemcee`, an adaptation of `emcee` (D. Foreman-Mackey et al. 2013) that incorporates dynamic temperature selection following W. D. Vousden et al. (2016).

#### 3.1.5. Best-fit Wavelength-independent Parameters

We modeled the combined NIRSpec white light curve using `batman` (L. Kreidberg 2015) for the planetary transit and `fleck` for the stellar surface. As the spot-crossing structure evolves between visits, we included spot rotation in this fit. To reduce dimensionality, we fixed the stellar rotation period to $P_{\rm rot} = 11.020$ days, from P. Tamburo et al. (2025), while using `ptemcee` to simultaneously fit for the stellar (inclination $i_*$, sky-projected obliquity $\lambda_*$, and quadratic limb-darkening coefficients $q_1$, $q_2$), orbital (planet-to-star radius ratio $R_{\rm p}/R_*$, period $P_{\rm orb}$, semimajor axis $a/R_*$, epoch $t_0$, and inclination $i$), spot (radius $R_{\rm spot}$, latitude $\phi_{\rm spot}$, longitude $\lambda_{\rm spot}$, and contrast $C$), and flare parameters. Priors and posterior values are summarized in Table 2.

As we normalized each visit before combining, we also included a constant scaling factor ($p_{2,0}$, $p_{4,0}$, $p_{5,0}$) for each visit that we multiply the transit model by to absorb offsets due to stellar rotation not captured by the normalization. Similarly to P. Tamburo et al. (2025), we restricted the stellar inclination to a uniform prior between 90° and 180°, while we allowed the sky-projected obliquity to vary from 0° to 360°, permitting the negative stellar pole to be anywhere on the visible hemisphere. This means that we allow the spot to rotate clockwise around the negative stellar pole for $i < 90°$, equivalent to a spot rotating counterclockwise around the positive pole with $i > 90°$ (P. Tamburo et al. 2025). We also included an uncertainty inflation parameter, $\sigma_{\rm infl}$, added in quadrature to the reported uncertainties to account for residual systematics and model incompleteness. Using the derived values of $i$, $i_*$, and $\lambda_*$, we calculate TOI-3884's true obliquity following Equation (4) from M. Mori et al. (2025).

We tuned the adaptation time and lag in `ptemcee` to 10 and 100, respectively. Sampling was performed with 25 temperatures and 50 walkers. After a burn-in of 80,000 steps, chains were run for an additional 5000 steps. All chains exceeded 50 times the integrated autocorrelation time, indicating convergence.[25]

### 3.2. *Joint Light-curve Analysis*

To construct a global spot model for TOI-3884, we performed a joint analysis of multiple JWST visits, allowing us to constrain the stellar inclination, sky-projected obliquity, rotation period, and spot parameters within a self-consistent rotating framework. We used `fleck`, which explicitly models spots on a rotating stellar surface.

#### 3.2.1. *Masking NIRISS Flares*

In addition to the three NIRSpec flares described in Section 3.1.2, the first two NIRISS visits exhibit large flare outburst events pre-ingress (see Figure 2). These events display complex, unusual morphologies (Figure 6 in Appendix B), making it difficult to identify individual flares or determine how many flare components to model. Additionally, the flare tails overlap with the transit and spot-crossing signal, further complicating joint modeling. Therefore, we excluded Visit 3 from the joint fit as the large flare outburst spans most of the transit (Figure 1), as indicated in the activity tracer H$\alpha$ (Figure 6 in Appendix B), preventing reliable separation of flare and spot signals. Visit 1 also shows flare activity; however, the H$\alpha$ emission dissipates throughout the spot-crossing. Therefore, we included Visit 1 but masked the flaring event.

#### 3.2.2. *White Light-curve Fit*

To model the joint white light curves, we fixed the wavelength-independent stellar and planetary parameters to the results from the NIRSpec joint fit (Table 2), while simultaneously fitting for the spot parameters ($R_{\rm spot}$, $\phi_{\rm spot}$, and $\lambda_{\rm spot}$), stellar rotation and geometry ($P_{\rm rot}$, $i_*$, and $\lambda_*$), and wavelength-dependent quantities: $R_{\rm p}/R_*$, $q_1$, $q_2$, $C$, $p$, and $\ln \sigma_{\rm infl}$. As in Section 3.1.5, each visit was normalized independently, so we included separate flux scaling parameters to absorb rotational flux variations introduced by `fleck`.

Previous long-baseline monitoring has measured rotation periods of $11.020 \pm 0.015$ days (P. Tamburo et al. 2025) and $11.043^{+0.054}_{-0.053}$ days (M. Mori et al. 2025). To reduce the complex covariances between stellar orientation, rotation, and spot parameters, we imposed a cautiously wide prior of 10–12 days on $P_{\rm rot}$.

[25] https://emcee.readthedocs.io/en/stable/tutorials/autocorr/

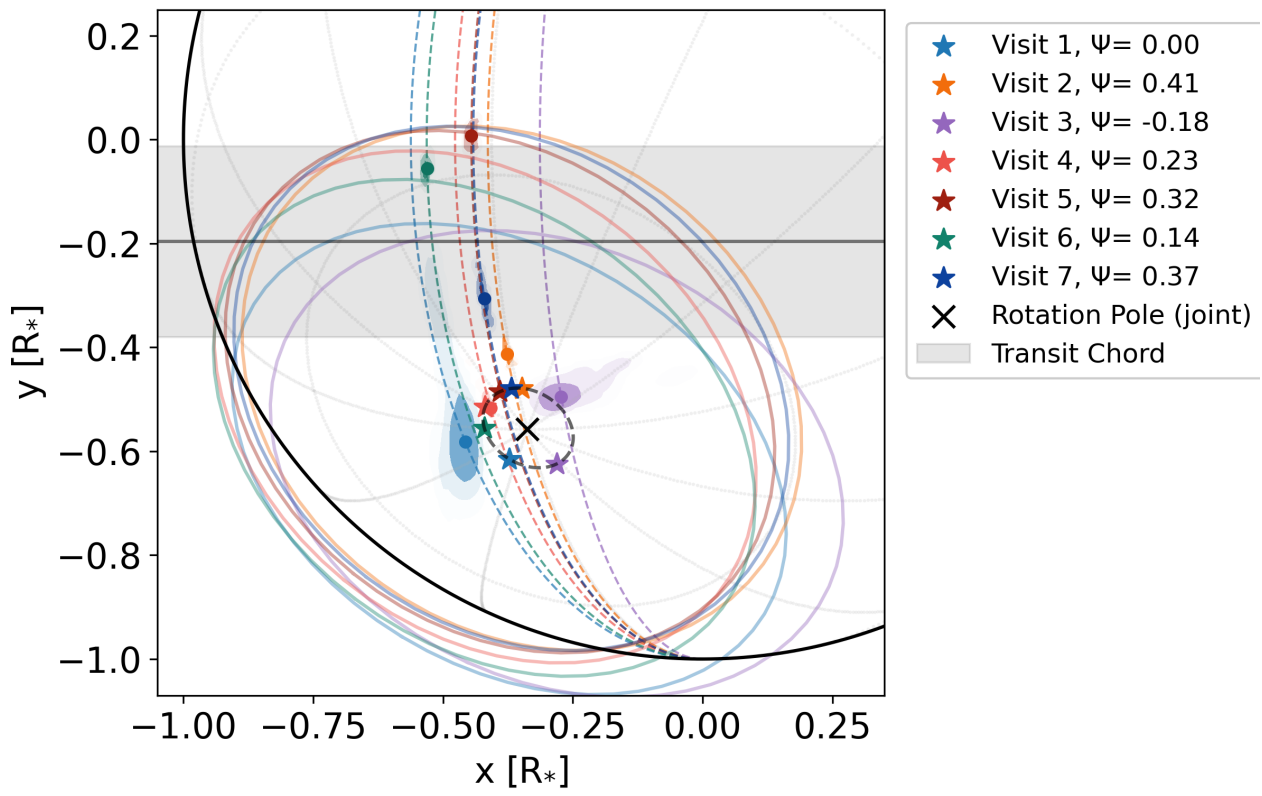

**Figure 3.** The best-fit TOI-3884 spot centers for the individual white light-curve fits (Section 3.3, solid circles with contours indicating the 1$\sigma$ and 2$\sigma$ regions from the posteriors) and the joint white light-curve fit (Section 3.2, stars). We plot ellipses indicating the extent of each spot from the self-consistent joint model. We find the best-fit solutions for Visits 5 and 6 have a spot above the transit chord (shaded gray), likely due to the geometric symmetry about the transit chord and the known radius–latitude degeneracy. Therefore, for those two visits, we also include the lines of projected "longitude" along which we might expect to find degenerate spots (dotted colored lines). The black cross marks the stellar rotation pole ($i_* = 139^{\circ}.2$, $\lambda_* = 31^{\circ}.1$) derived in the self-consistent joint model, and the dashed black line indicates how the center of the spot rotates around the pole. The stellar rotation phase, $\Psi$, for each visit is shown in the legend (assuming $\Psi = 0$ at the first visit). This plot is inspired by Figure 3 from H. Chakraborty et al. (2025).

Posterior sampling was performed using `ptemcee` with 25 temperatures, 50 walkers, and a burn-in of 18,000 steps followed by 1000 production steps. The resulting best-fit rotating spot model is shown in Figure 2. Derived spot locations are presented in Figure 3, and the spot contrasts and radii are shown in Figure 4. The joint-fit parameter inferences are summarized in the last column of Table 2.

We note that the large uncertainty inflation parameter (Table 2) for NIRISS white light curves ($\ln \sigma_{\rm NIRISS} = 2.51$; $\sigma_{\rm NIRISS} \approx 12$). The NIRISS uncertainties are likely significantly underestimated, as we do not account for the structured noise in the residuals (as seen in Figure 2). Future work accounting for flares, multiple spots, and additional spot complexity should better model the remaining astrophysical noise in the residuals.

#### 3.2.3. *Spectroscopic Light Curves*

To extract the panchromatic spot contrast spectra, we first binned Visits 1, 2, 4, 5, 6, and 7 to a spectral resolving power of $R = 100$. We assumed the spot geometry ($R_{\rm spot}$, $\phi_{\rm spot}$, $\lambda_{\rm spot}$) does not vary with wavelength and adopted the white-light best-fit values for each visit (Section 3.2.2). As for the white light-curve fits, we mask all flares except for those in Visit 5.

We then modeled the $R = 100$ spectroscopic light curves for each visit separately, fitting for the wavelength-dependent variables ($R_{\rm p}/R_*$, $q_1$, $q_2$, $C$, $p_0$, and $\ln \sigma_{\rm infl}$), but now fixing the stellar orientation ($i_* = 139^{\circ}.2 \pm 0^{\circ}.3$, $\lambda_* = 31^{\circ}.1 \pm 0^{\circ}.4$), rotation period ($P_{\rm rot} = 11.102 \pm 0.003$ days), spot radius ($R_{\rm spot}$=0.576 $^{+0.006}_{-0.005}$ $R_*$), and spot locations according to the joint white-light solution. Only wavelength-dependent parameters were allowed to vary. The resulting contrast spectra for all visits (except Visit 1) are displayed in Figure 5. Though the spectra from Visit 1 are consistent, the uncertainties due to significant masking are much larger; therefore, for clarity, we

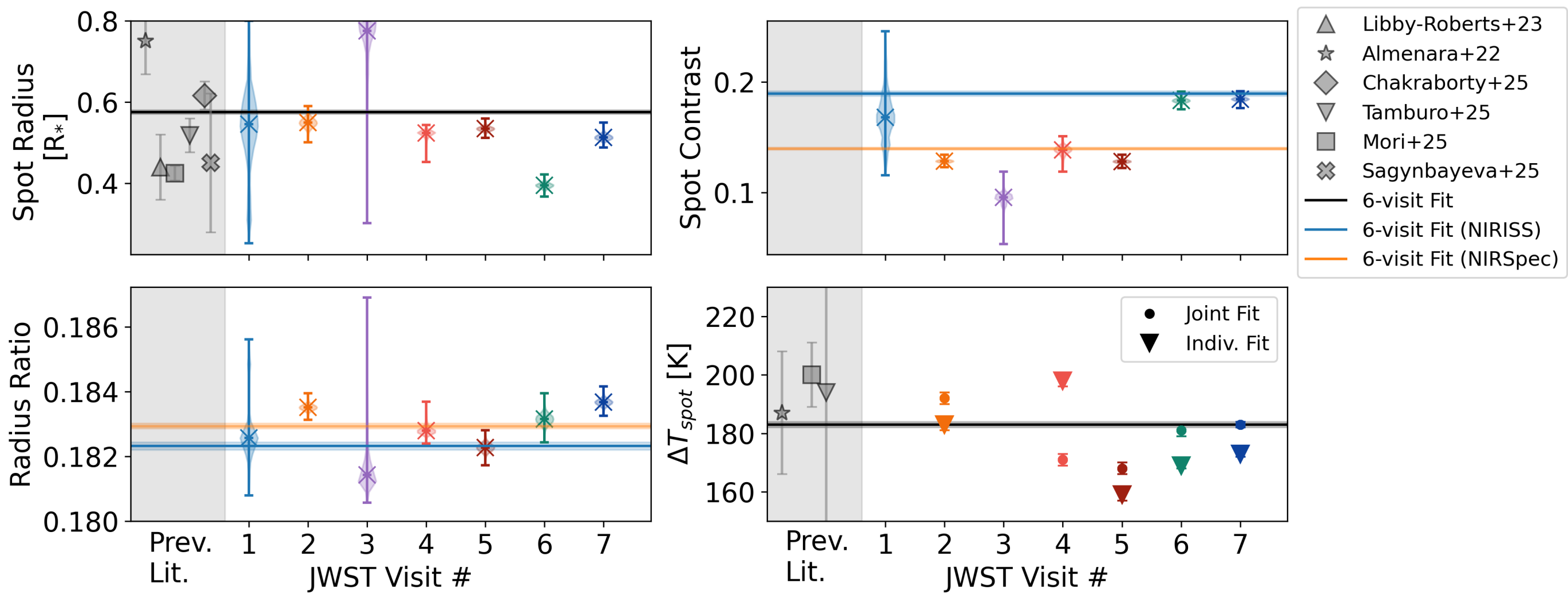

**Figure 4.** Best-fit TOI-3884 spot radii (top left), instrument-dependent contrasts (upper right), and planet-to-star radius ratios (lower left) from each individual white light-curve fit. The best-fit spot radius, contrasts, and radius ratios from the self-consistent joint model are indicated with horizontal lines. Visits 1 and 3 are significantly impacted by stellar flares, therefore are heavily masked, inflating the uncertainties of the derived parameters for these visits. The best-fit spot temperature for Visits 2, 4, 5, 6, and 7 are shown on the bottom right plot. For comparison, we include literature values for the wavelength-independent spot radius and temperature ($187 \pm 21$ K, J. M. Almenara et al. 2022; $194 \pm 72$ K, P. Tamburo et al. 2025; and $200 \pm 11$ K, M. Mori et al. 2025). We note that J. E. Libby-Roberts et al. 2023 report a large temperature range ($\Delta T \approx 300$–$500$ K) out of plot range, while H. Chakraborty et al. (2025) and S. Sagynbayeva et al. (2025) do not quote temperatures.

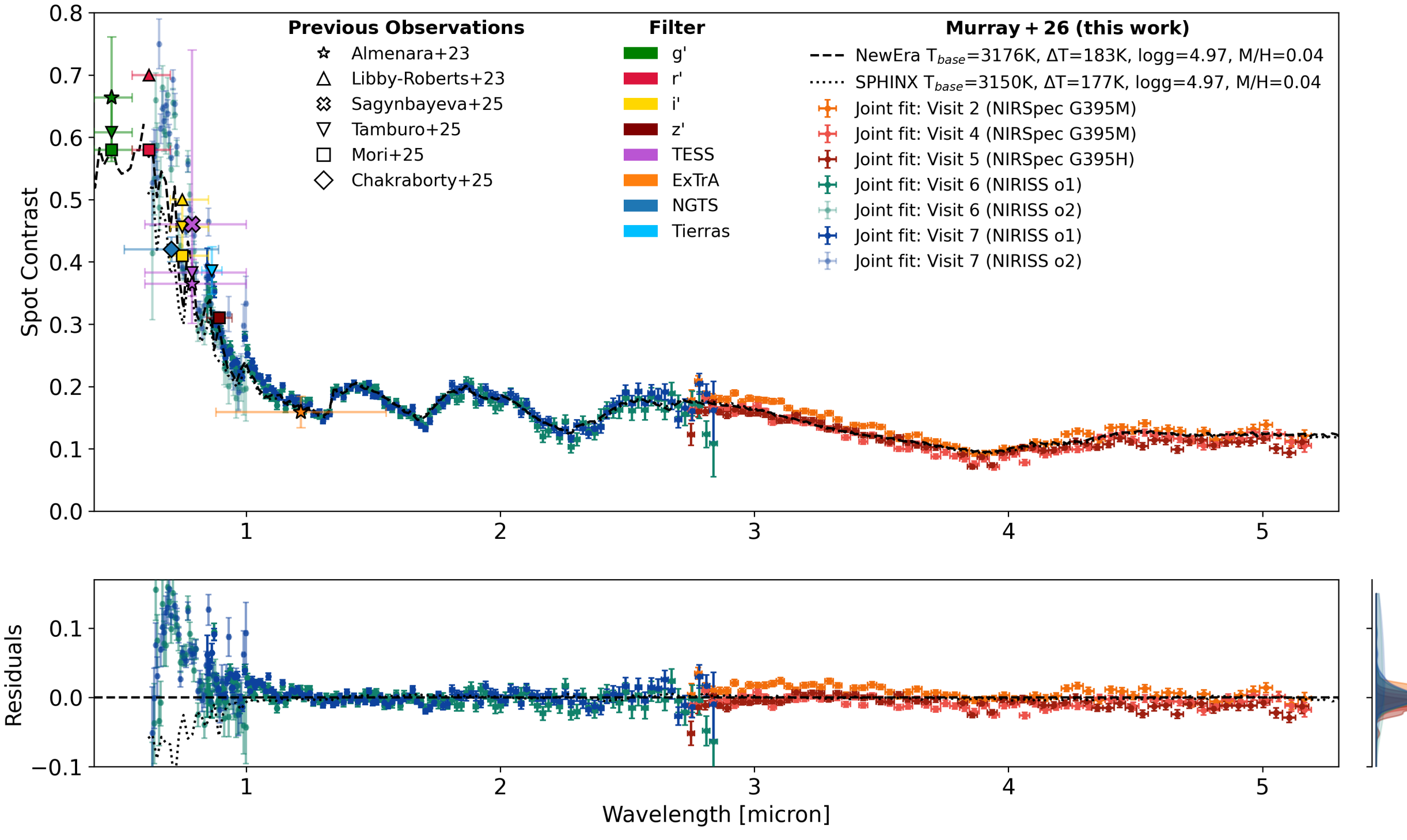

**Figure 5.** Top: the empirical spot contrast spectra for JWST Visits 2, 4, 5, 6, and 7. The contrast uncertainties are those derived directly from the spectroscopic fits in Section 3.2.3. The dashed black line shows the best-fit NewEra model ($T_{\rm base} = 3176 \pm 10$ K, $T_{\rm spot} = 2993 \pm 10$ K) and the dotted line shows the best-fit SPHINX model ($T_{\rm base} = 3150 \pm 13$ K, $T_{\rm spot} = 2972 \pm 13$ K). We overplot broadband contrasts derived in previous works by J. M. Almenara et al. (2022), J. E. Libby-Roberts et al. (2023), P. Tamburo et al. (2025), M. Mori et al. (2025), S. Sagynbayeva et al. (2025), and H. Chakraborty et al. (2025) (compiled in Table 4). Bottom: residuals between the spectroscopic contrasts and the best-fit NewEra model. The difference between the SPHINX and NewEra models is shown with the dotted black line. A histogram of the residuals for each visit is shown to the right of the residual plot. The JWST spectra shown in the top plot are available as the data behind the figure in machine-readable format.

(The data used to create this figure are available in the online article.)

opted to omit them from Figures 5 and 7. The transmission spectrum, derived from $R_{\rm p}/R_{*}$, will be presented in future work.

### 3.3. Individual Light-curve Analysis

As a cross-check for our self-consistent joint spot model in Section 3.2, we also performed the same fit on each JWST visit individually. This method also allows us to quantify the potential influence of imperfect spot geometry. Although the giant polar spot appears to be long-lived (J. M. Almenara et al. 2022), the 6 month separation between the GO-5863 and GO-5799 observations raises the possibility of surface evolution. Additionally, our model assumes a single circular spot, whereas the true morphology may consist of multiple smaller spots, noncircular structures, or regions with varying temperatures. Fitting each visit separately, therefore, allows us to characterize the stellar surface at each epoch without imposing constraints from stellar rotation or long-term stability.

We provide full details of the individual visit fits in Appendix C, with the resulting best-fit spot models for each visit illustrated in Figure 2, locations shown in Figure 3, contrasts and radii in Figure 4, and spot contrast spectra presented in Figure 7 in Appendix C.2. Though we derive contrast spectra for each visit, we primarily perform this analysis for robustness, and so we recommend that readers use the results from the joint visit model (spot parameters from the last column of Table 2 and the spectrum from Figure 5).

## 4. The Polar Spot Spectrum

From the spectroscopic fits for both the joint and individual visit models, we measure the spot contrast as a function of wavelength. The resulting spectra are shown in Figures 5 and 7.

We define spot contrast as

$$C(\lambda) = 1 - \frac{S_{\rm spot}(T_{\rm spot}, \log g, Z, \lambda)}{S_{\rm base}(T_{\rm base}, \log g, Z, \lambda)}, \quad (1)$$

where $S_{\rm spot}$ and $S_{\rm base}$ are the emergent spectra of the spotted and quiescent (baseline) photosphere, respectively, characterized by temperatures $T_{\rm spot}$ and $T_{\rm base}$, surface gravity $\log g$, and metallicity, $Z$. We assume that $\log g$ and $Z$ are the same for the spot and the photosphere, although recent work has explored fitting these parameters separately (M. Fournier-Tondreau et al. 2024). Using this formalism, $C = 0$ corresponds to a spot with the same temperature as the quiet photosphere, while $C = 1$ represents a perfectly dark spot ($T_{\rm spot} = 0$ K); therefore, larger temperature differences correspond to larger spot contrasts. We note that `fleck` defines spot contrast inversely, such that a completely dark spot has a contrast of zero.

### 4.1. Comparison with 1D Stellar Models

We used `speclib`[26] (B. V. Rackham 2023; B. V. Rackham & J. de Wit 2024) to extract and interpolate between NewEra (P. H. Hauschildt et al. 2025) and low-resolution SPHINX (A. R. Iyer et al. 2023) stellar atmosphere grids spanning $2300 \leqslant T_{\rm eff} \leqslant 4000$ K. Assuming the spot can be approximated as a cooler photospheric component, we also extracted NewEra/SPHINX models for temperatures cooler than the photosphere $0 \leqslant \Delta T \leqslant 500$ K. We fixed $\log g = 4.97$ dex and $Z = 0.04$ following J. E. Libby-Roberts et al. (2023), though we later explore the results of relaxing this assumption.

We fit the NewEra models to our five contrast spectra from Section 3.2.3, sharing $T_{\rm base}$, $T_{\rm spot}$, $\log g$, and $Z$ across all visits. Because the bluer NIRISS contrasts ($<1\ \mu$m) deviate significantly from the 1D model predictions (Figure 5), we restricted the fits to wavelengths longer than 1 $\mu$m. We also fit two additive jitter parameters: one for NIRISS and one for NIRSpec wavelengths. Posterior sampling was performed with `emcee` using 50 walkers, 10,000 burn-in steps, and 5000 production steps, verifying convergence via the autocorrelation time.

**Table 3**
Spot Contrast Temperatures, $\Delta T$, and Quiescent Photospheric Temperatures, $T_{\rm base}$, from Best-fit NewEra Models When (a) the Spot Temperature is Constant across All Visits, or (b) the Spot Temperature is Independently Fit for Each Visit

| | (a) Shared $\Delta$T | |
|---|---|---|
| | Individual | Joint |
| $T_{\rm base}$ | 3111 ± 13 K | 3176 ± 10 K |
| $\Delta T$ | 170 ± 1 K | 183 ± 1 K |
| | **(b) Independent $\Delta$Ts** | |
| $T_{\rm base}$ | 3146 ± 12 K | 3150 ± 12 K |
| $\Delta T_2$ | 183 ± 2 K | 192 ± 2 K |
| $\Delta T_4$ | 198 ± 2 K | 170 ± 2 K |
| $\Delta T_{5,1}$ | 166 ± 2 K | 176 ± 2 K |
| $\Delta T_{5,2}$ | 151 ±3 K | 160 ± 3 K |
| $\Delta T_6$ | 169 ± 1 K | 181 ± 1 K |
| $\Delta T_7$ | 173 ± 1 K | 183 ± 1 K |
| $\overline{\Delta T}$ | 171 ± 7 K | 179 ± 5 K |

We fit the contrast spectra for two different scenarios: (1) we assume a single spot temperature across all visits, and (2) we allow the spot temperature to vary between visits. For scenario (1), we derive $T_{\rm base}$=3176 ± 10 K and $\Delta T_{\rm spot} = T_{\rm base} - T_{\rm spot} = 183 \pm 1$ K. We include this best-fit NewEra model in Figures 5 and 7. For scenario (2), the results are shown in Table 3 and Figure 4. Though visit-by-visit spot temperatures vary, on average, we find consistency between (a) shared and (b) independent $\Delta T$ assumptions, and also between the individual (Section 3.3) and joint (Section 3.2) spot models.

We verified our results by repeating (a) with SPHINX models, yielding $T_{\rm base} = 3149 \pm 13$ K and $\Delta T_{\rm spot}$=177 ± 2 K, consistent with NewEra. We overplot this model in Figures 5 and 7, additionally showing the residuals between NewEra and SPHINX. Although both models underestimate the observed spot contrasts, NewEra better reproduces the contrasts for $\lambda \lesssim 1\ \mu$m (Figure 5; residuals up to ~15% with NewEra compared to ~20% with SPHINX, and the SPHINX fit requires more additive jitter at NIRISS wavelengths than NewEra).

In these fits, we fixed the stellar metallicity and gravity. When allowed to vary, the sampling pushes against the upper prior limits of $\log g = 6.0$ and $Z = +0.5$, suggesting that the 1D models attempt to compensate for mismatches in the contrast spectrum. Since fitting the contrast spectra is highly dependent on the quality of the spot fitting and the agreement with 1D stellar models, we do not consider these results reliable. Mass and radius estimates from empirical relations (A. W. Mann et al. 2015, 2019) predict $\log g$=4.920 ± 0.027 dex, in line with the adopted value of $\log g = 4.97$ dex. As J. E. Libby-Roberts et al. (2023) derived TOI-3884's properties using high-resolution spectra from Habitable-zone Planet Finder (S. Mahadevan et al. 2012, 2014), we adopt those values here. In future work, we will perform a full out-of-transit spectral energy distribution (SED) analysis of the JWST data to reassess the stellar properties and evaluate the impact of the giant polar spot on these parameters.

[26] https://github.com/brackham/speclib

## 5. Discussion

TOI-3884 b is a super-Neptune planet transiting over the pole of an M dwarf that hosts a giant polar spot. By observing this unusual system with JWST and fitting the persistent spot-crossing events in transit, we derive a benchmark panchromatic spot contrast spectrum from 0.6 to 5.3 $\mu$m.

### 5.1. Individual versus Joint Visit Fits

Our individual white light-curve results (see Figures 2, 3, and 4) favor a spot that is smaller and closer to the transit chord than our self-consistent joint model. For the individual fits, we allow the spot to be any size and to exist anywhere on the stellar surface. For the joint model, we constrain the spot properties by asserting that we have one spot (with constant radius and wavelength-dependent contrast) that is rotating about a fixed pole.

From Figure 2, we can see that the individual spot fits often describe the data better than our joint model. As we allow the spot full flexibility in the individual fits, this is to be expected when there is additional complexity beyond a single stable circular starspot. The limitations of the joint model are most evident for Visit 3, which is dominated by a flare outburst, as it does a poor job of replicating the light-curve structure. Visit 3 is also the only dataset sampling the opposite half of the stellar rotation phase ($-0.5 < \Psi < 0$), whereas the remaining visits lie at $0 < \Psi < 0.5$ (Figure 3); removing it therefore weakens constraints on the global geometry in Section 3.2. Improving the global model will likely require additional phase coverage uncontaminated by large flares and/or more flexible spot prescriptions (e.g., multiple spots or structured active regions). Nevertheless, the agreement between the contrast spectra inferred from the individual and joint analyses (Figures 5 and 7) shows that the joint framework is sufficient for the primary goal of this work: establishing an empirical spot spectrum.

Despite differences in the spot-crossing profiles, we find consistent contrasts for the spectroscopic fits between the joint visit and individual visit models. This reflects the fact that the contrast is primarily set by the amplitude of the spot-crossing anomaly, which is less sensitive to the radius–latitude degeneracy than the spot location itself. The largest discrepancies occur in Visits 1 and 3, where strong flare activity likely blends into the spot-crossing morphology even after masking.

Figure 7 in Appendix C.2 shows that the individually fit spot contrast spectra agree well at NIRISS wavelengths ($\Delta T_{\rm spot} =$ 169 $\pm$ 1 K and 173 $\pm$ 1 K for Visits 6 and 7), but there are larger differences among the three NIRSpec visits ($\Delta T_{\rm spot} =$ 183 $\pm$ 2 K, 198 $\pm$ 2 K, 166 $\pm$ 2 K, and 151 $\pm$ 3 K for Visits 2, 4, and 5 NRS1 and NRS2—a 47 K range). These differences do not appear to be correlated with time, implying they are not due to the evolution of the giant spot. One possible explanation is astrophysical: although the intrinsic spot does not change, TOI-3884 b occults a different mixture of penumbra, umbra, and faculae at different stellar rotation phases that lead to apparent variations in the spot contrast. Alternatively, these differences may arise from the fitting process itself, as independently fitting each visit allows for slightly different solutions within the degenerate radius–latitude–contrast parameter space. Consistent with this interpretation, the self-consistent joint model reduces visit-to-visit scatter in the redder contrasts (a 32 K range; see Figure 5), implying that at least part of the variation seen in the individual fits is likely due to degeneracy-driven difficulties in the inferred spot geometry.

We also caveat that the small uncertainties on our spot temperature ($\pm$1 K) are statistical error bars from MCMC sampling, which may not fully capture the systematic uncertainties from the mismatch with stellar models, particularly at the short wavelengths we removed when fitting.

### 5.2. Flares

TESS photometry of TOI-3884 shows only a few definitive flares across three sectors (J. E. Libby-Roberts et al. 2023). In our JWST dataset—taken over 2 yr later—we detect multiple small flare events, along with two large flare outbursts in Visits 1 and 3. This difference could reflect a higher activity state at the end of 2024, or it may also result from JWST's higher precision, which enables detection of weaker flares despite the redder bandpass and decreased contrasts. Contemporaneous ground-based observations also detect flares near our JWST epochs: two flares on 2024 December 11 and 2025 January 4 with Las Cumbres Observatory (M. Mori et al. 2025), and one flare on 2025 May 22 with Tierras (P. Tamburo et al. 2025).

Notably, the two large flare outbursts in Visits 1 and 3 occur almost exactly at the same orbital phase, shortly before ingress. TOI-3884 b orbits just within the star's Alfvén radius, where magnetic star–planet interactions (SPI) can, in principle, propagate energy back to the solar surface (S. Preusse et al. 2006; O. Cohen et al. 2011). One possibility is that the planet's passage near the polar active region perturbs the strong magnetic field there and triggers enhanced flare activity, consistent with SPI scenarios (E. Ilin et al. 2023, 2025; N. Whitsett & T. Daylan 2025). Long-term photometric monitoring will be needed to constrain the star's flaring rate and ascertain whether these events were coincidental or the result of SPI. Additionally, these processes can induce radio emission, offering an independent diagnostic (e.g., J. S. Pineda & J. Villadsen 2023; K. N. O. Ceballos et al. 2025; L. Peña-Moñino & M. Pérez-Torres 2025).

### 5.3. Limb Darkening

We test whether our choice of quadratic limb darkening (the only option in `fleck`) affects our results by comparing quadratic, three-parameter, and four-parameter laws using ExoTiC-LD (D. Grant & H. R. Wakeford 2024). For our $R = 100$ spectroscopic light curves, all parameterizations yield consistent transit profiles within the data uncertainties. We further test the NIRISS data, where limb darkening is strongest, by masking the spot-crossing and fitting quadratic and three-parameter laws using BATMAN. We find no significant differences in recovered transit depths or spectroscopic light curves. In both cases, fitted coefficients deviate from theoretical predictions, likely reflecting either limitations in stellar atmosphere models or residual systematics associated with unmodeled spot complexity (see Figure 2).

Finally, we assess potential biases from the Kipping parameterization in the weak limb-darkening regime using the NIRSpec/G395H data, following L.-P. Coulombe et al. (2024). Fitting the egress-only light curve with quadratic limb darkening and broad uniform priors ([−3, 3]) yields limb-darkening coefficients and transit depths consistent with the results from Sections 3.2 and 3.3 within uncertainties. We therefore find no evidence that our results are biased by the choice of limb-darkening law or parameterization. A full exploration of limb darkening, including comparison to 3D MHD models and more complex spot geometries, is left to future work.

### *5.4. Spot Modeling Limitations*

Our spot modeling uses `fleck`. As noted in Section 3.1.3 and as can be seen in Figure 3, `fleck` can produce "overhanging" spots near the stellar limb; if such an overhang intersects the transit chord, it can create bright artifacts in the light curve before or after transit. We see no (or minimal) overhang in the transit chord (Figures 2 and 3), implying negligible impact on the inferred spot-crossing shape. However, this behavior in `fleck` may still affect the likelihood landscape by disfavoring some spot solutions and potentially biasing the preferred spot position away from the mid-transit point. Reassuringly, we do note that our spot position with respect to the planet is broadly consistent by eye with analyses using different tools, including `SAGE` (H. Chakraborty et al. 2025) and `starry` (P. Tamburo et al. 2025).

As an independent check, we also compared our best-fit parameters from `fleck` with `spotter`[27] (L. Garcia et al. 2025), which models surface features with a pixelated stellar map. Fixing the stellar inclination and obliquity to the joint-fit values, we performed maximum a posteriori optimization (without full posterior sampling) using `spotter` for five JWST visits (2, 4, 5, 6, 7) individually. We found spot radii ranging from 0.55 to 0.65 $R_*$, consistent with our best-fit radius of 0.576 $^{+0.006}_{-0.005}$ $R_*$, and with an average difference of 0.03 $R_*$ between the `spotter` and `fleck` radii. The spot latitudes from `spotter` are all $>77°$, with an average difference of 5° from `fleck` ($-84\overset{\circ}{.}69 \pm 0\overset{\circ}{.}12$). The `spotter` spot contrasts ($C_{\rm NIRISS} = 0.27$ and $C_{\rm NIRSpec} = 0.18$) are on average 0.05 larger than the `fleck` values; however, given the lack of full posterior sampling and the agreement of our contrast spectrum with previous results, we do not consider this offset significant.

Finally, our analysis neglects second-order effects from differential limb darkening across the spotted versus the unspotted photosphere. As this spot is both large and on the limb, such effects could be nonnegligible. Quantifying them, however, likely requires detailed 3D MHD modeling (e.g., N. Kostogryz et al. 2023; N. M. Kostogryz et al. 2024) of fully convective M dwarfs like TOI-3884, which is beyond the scope of this Letter.

### *5.5. How Well Do Models Fit the Observed Contrasts?*

Despite known discrepancies between 1D models and M dwarf spectra, our contrast spectra at wavelengths $>1\ \mu$m agree surprisingly well with current NewEra models. We fit for additive jitter terms for NIRISS order 1 and NIRSpec to capture some model discrepancies, which we find to be $\sigma_{\rm NIRISS} = 0.0070 \pm 0.0007$ and $\sigma_{\rm NIRSpec} = 0.0099 \pm 0.0007$. If we perform the same fit but include NIRISS data $<1\ \mu$m, we find the jitter term for NIRISS order 2 wavelengths (0.6–1 $\mu$m) is 5–6× that of NIRISS order 1 and NIRSpec.

At these blue wavelengths, this may indicate missing opacities from dominant sources in M dwarf atmospheres, such as the metal oxides TiO and VO (A. S. Rajpurohit et al. 2013; A. R. Iyer et al. 2023). These discrepancies between our spot contrast spectrum and stellar models imply that if we were to fit only optical-wavelength contrasts (without near-IR information), we would likely extract different spot and stellar properties. Numerous other works have found differences in the stellar properties derived from optical and infrared data for M dwarfs (e.g., T. Olander et al. 2021, Figure 6 from V. M. Passegger et al. 2016, and Figures 5 and 6 from D. J. Wilson et al. 2025). We also find that the size of the FeH feature at 1 $\mu$m is underestimated by NewEra models, potentially due to difficulties modeling pressure broadening for low-mass stars (A. Glidden et al. 2026). Full interpretation of these discrepancies is beyond the scope of this Letter, but our empirical spectrum provides a direct benchmark for testing next-generation 3D MHD predictions (e.g., V. Witzke et al. 2024; H. N. Smitha et al. 2025) in future work.

### *5.6. Stellar Contamination of Transmission Spectra*

Our empirical spot contrasts in the NIRISS order 2 bandpass ($\lambda = 0.6$–$1\ \mu$m) show poor agreement with the 1D stellar models typically used for TLS corrections. Therefore, it is possible that some signatures detected at short wavelengths on transiting planets, interpreted as Rayleigh scattering from the atmospheric gas or hazes, could be due to stellar contamination not properly corrected at short wavelengths, even when anchored well at redder wavelengths. As such, we recommend caution in interpreting signals at $\lambda < 1\ \mu$m in NIRISS transmission spectra, even if a TLS correction has been performed.

However, we find that in the NIRSpec bandpass ($\lambda = 3$–$5\ \mu$m) the stellar contamination signal is smoothly varying and agrees well with 1D model expectations. This suggests that contamination is unlikely to dominate molecular detections for prominent absorbers—such as $CH_4$, $CO_2$, or $SO_2$—in this wavelength range. More generally, these results highlight the advantage of NIRSpec observations in constraining planetary atmospheres in the presence of stellar activity, especially for systems with small planets transiting low-mass M dwarfs like TRAPPIST-1.

### *5.7. Comparison with Previous TOI-3884 b Studies*

Since its discovery, TOI-3884 b has been observed by many ground-based telescopes, including ExTrA (J. M. Almenara et al. 2022), Apache Point Observatory (APO; J. E. Libby-Roberts et al. 2023), Tierras (P. Tamburo et al. 2025), MuSCAT (M. Mori et al. 2025), and Euler (H. Chakraborty et al. 2025). Our JWST observations extend this record by providing seven precisely measured transits over 5 months, all showing a spot-crossing anomaly during the first half of the transit (Figure 2). This confirms the presence of a large, stable high-latitude active region and corroborates prior reports of visit-to-visit changes in the detailed spot morphology (J. E. Libby-Roberts et al. 2023; H. Chakraborty et al. 2025; M. Mori et al. 2025; P. Tamburo et al. 2025).

By fitting six JWST visits jointly with a single rotating spot model (Section 3.2), we constrain the stellar orientation and rotation period. We determine the stellar inclination to be $i_* = 139\overset{\circ}{.}2 \pm 0\overset{\circ}{.}3$, sky-projected obliquity to be

[27] https://spotter.readthedocs.io/en/latest/

$\lambda_* = 31\overset{\circ}{.}1 \pm 0\overset{\circ}{.}4$, and true obliquity to be $\psi_* = 56\overset{\circ}{.}4 \pm 0\overset{\circ}{.}3$, consistent with a misaligned orientation in which TOI-3884's pole is directed either directly toward or away from us and the planet occults a high-latitude spotted region. While stellar inclinations are dependent on the direction of the stellar pole, our true obliquity of $\psi_* = 56\overset{\circ}{.}4 \pm 0\overset{\circ}{.}3$ is consistent with literature values of $50° \pm 12°$ (J. M. Almenara et al. 2022), $61.9^{+2.3}_{-5.6}$deg (M. Mori et al. 2025), and $54\overset{\circ}{.}3 \pm 1\overset{\circ}{.}4$ (H. Chakraborty et al. 2025). This orientation naturally explains why the large spot-crossing signatures do not correspond to strong rotational modulation in TESS or ground-based photometry.

Long-baseline monitoring yields consistent stellar rotation periods of $11.043^{+0.054}_{-0.053}$ days (M. Mori et al. 2025) and $11.020 \pm 0.015$ days (P. Tamburo et al. 2025). Our rotation period estimate agrees with M. Mori et al. (2025) within $1.1\sigma$ (though notably the value in M. Mori et al. 2025 has a larger uncertainty), however is slightly discrepant with P. Tamburo et al. (2025) at the $5.5\sigma$ level. Although our $P_{\rm rot}$ appears more precise (probably due to sharp multimodal posterior distributions), the $P_{\rm rot}$ values from M. Mori et al. (2025) and P. Tamburo et al. (2025), which are in agreement with each other, are likely more robust, as they rely only on periodic brightness variations rather than assumptions about spot shape, stability, and multiplicity. Though we implemented a 10–12 days prior, it is reassuring that our estimate for $P_{\rm rot} = 11.102 \pm 0.003$ days, derived from modeling the time-dependent shape of the spot crossings in the light curves, aligns fairly well with independent derivations of $P_{\rm rot}$ from tracking the variation in TOI-3884's brightness. It is also possible that our derived rotation period is slower than that of M. Mori et al. (2025) and P. Tamburo et al. (2025) due to differential rotation: we model the rotation of a polar spot compared to stellar disk-averaged brightness fluctuations. This $\Delta P_{\rm rot} \approx 0.06$–$0.08$ day would correspond to a differential shear between pole and average disk of $\Delta\Omega \approx 0.003$–$0.004$ rad day$^{-1}$ (representing a lower limit on equator-to-pole differential rotation) that is consistent with other M dwarfs such as GJ 1243 (see Figure 7 of J. R. A. Davenport et al. 2015).

Using the three NIRSpec visits, we jointly constrain the orbital parameters of $a/R_* = 24.73 \pm 0.04$, $P = 4.5445836 \pm 0.0000005$ days, $i = 89\overset{\circ}{.}54 \pm 0\overset{\circ}{.}02$, and $b = 0.199 \pm 0.009\,R_*$. Our $a/R_*$ value is consistent with that of J. M. Almenara et al. (2022) and P. Tamburo et al. (2025), who find $a/R_*$ values of $25.01 \pm 0.65$ and $25.06 \pm 0.91$, and within $2\sigma$ of J. E. Libby-Roberts et al. (2023)'s value of $25.9 \pm 0.9$. In contrast, our value is lower by $10\sigma$ relative to H. Chakraborty et al. (2025)'s value of $25.7 \pm 0.1$, while all of these values are notably higher than M. Mori et al. (2025)'s value of $23.5 \pm 0.2$. One plausible contributor is the strong degeneracy between $i$ and $a/R_*$ through the transit duration, which is difficult to constrain when limb features—such as TOI-3884 b's spot-crossing anomaly—distort ingress and egress.

Assuming a circular orbit, our results imply a stellar density of $\rho_* = 13.85 \pm 0.07$ g cm$^{-3}$. This is consistent with $\rho_* = 15.3 \pm 2.0$ g cm$^{-3}$ from the SED/isochrone analysis of J. E. Libby-Roberts et al. (2023). Using empirical relations (A. W. Mann et al. 2015, 2019), we also infer $M_* = 0.282 \pm 0.006\,M_\odot$ and $R_* = 0.305 \pm 0.009\,R_\odot$, corresponding to $\rho_* = 14.1 \pm 1.3$ g cm$^{-3}$, again consistent with our result.

**Table 4**
Broadband Spot Contrast Values from Previous TOI-3884 b Studies

| References | Filter(s) | $C$ |
|---|---|---|
| J. M. Almenara et al. (2022) | $g'$[a] | $0.664 \pm 0.097$ |
| | TESS[b] | $0.365^{+0.064}_{-0.039}$ |
| | ExTrA[c] | $0.159 \pm 0.025$ |
| J. E. Libby-Roberts et al. (2023) | $r'$[a] | 0.5 |
| | $i'$[a] | 0.5 |
| M. Mori et al. (2025) | $g'$ | 0.58 |
| (and private communication) | $r'$ | 0.58 |
| | $i'$ | 0.41 |
| | $z'$[a] | 0.31 |
| P. Tamburo et al. (2025) | $g'$ | $0.608^{+0.047}_{-0.052}$ |
| | $i'$ | $0.456^{+0.043}_{-0.044}$ |
| | TESS | $0.383 \pm 0.038$ |
| | Tierras[d] | $0.386 \pm 0.038$ |
| H. Chakraborty et al. (2025) | NGTS[e] | $0.42 \pm 0.02$ |
| S. Sagynbayeva et al. (2025) | TESS | $0.46^{+0.28}_{-0.1}$ |

**Notes**
[a] The $g'$, $r'$, $i'$, and $z'$ observations were taken with the Sloan/SDSS filters.
[b] $\lambda = 0.6$–$1.0\ \mu$m with central wavelength $\lambda c = 0.7865\ \mu$m.
[c] $\lambda = 0.88$–$1.55\ \mu$m; $\lambda c = 1.215\ \mu$m.
[d] $\lambda = 0.8235$–$0.9035\ \mu$m; $\lambda c = 0.8635\ \mu$m (J. Garcia-Mejia et al. 2020).
[e] $\lambda = 0.520$–$0.890\ \mu$m; $\lambda_c = 0.705\ \mu$m (P. J. Wheatley et al. 2018).

From fitting the spot contrast spectrum, we find that TOI-3884's starspot has a temperature $183 \pm 1$ K cooler than the quiescent photosphere temperature of $3176 \pm 10$ K. Spot temperature contrasts of order ~200 K are expected for mid M dwarfs (S. V. Berdyugina 2005), in alignment with our values. Figure 5 and Table 4 also include derived broadband contrast values from a number of previous photometric TOI-3884 b studies. We find that our contrast spectrum, extended contrast model, and derived temperatures agree reasonably well with the literature. Additionally, the residuals between the photometry and the best-fit NewEra model in Figure 5 appear to show a similar offset to our spectrum at bluer wavelengths ($<1\ \mu$m). Agreement is strongest with studies using redder filters, consistent with the growing mismatch between empirical contrasts and 1D models at $\lambda < 1\ \mu$m, which would drive higher inferred contrasts if interpreted through those models. Overall, our JWST-based spectrum provides the first continuous, panchromatic measurement of the polar spot contrast from 0.6 to 5.3 $\mu$m, placing these broadband measurements in a unified spectral context.

## 6. Conclusions

We have presented an analysis of seven JWST transit observations of TOI-3884 b obtained with NIRISS/SOSS and NIRSpec/G395M and G395H, spanning 0.6–5.3 $\mu$m. These observations repeatedly capture the planet occulting a large stellar active region on its M dwarf host star. By modeling both white-light and spectroscopic time series data across multiple epochs, we use TOI-3884 b as a probe of the stellar surface, constraining the system geometry, the properties of the persistent polar spot, and the wavelength-dependent contrast between the spot and the surrounding photosphere. This unique dataset enables the first empirical, panchromatic (0.6–5.3 $\mu$m) JWST spectrum of a starspot on an M dwarf, providing a critical empirical benchmark for stellar atmosphere models and contamination corrections.

Our main conclusions are as follows:

1. We confirm that TOI-3884 hosts a large, long-lived polar active region that is occulted in every observed transit, spanning 5 months. All seven JWST visits exhibit a pronounced spot-crossing anomaly in the transit light curves. While the detailed morphology of the anomaly evolves between visits, its consistent recurrence indicates that the spot structure is stable on timescales much longer than the planetary orbital period.
2. Stellar flares are frequent but do not obscure the underlying spot-crossing signal. All visits show evidence of stellar flaring to some degree, with particularly strong and complex flare activity in the earliest NIRISS observations. Despite this activity, the spot-crossing signal remains clearly detectable in every transit. While visits affected by strong flaring activity exhibit larger residuals and broader posteriors in individual fits, joint multi-visit modeling yields consistent stellar orientation and spot properties across the full dataset.
3. Repeated spot crossings tightly constrain the stellar rotation period and three-dimensional stellar orientation. By modeling the timing and morphology of repeated spot-crossing events across multiple epochs, we measure a stellar rotation period of $P = 11.102 \pm 0.003$ days and constrain the stellar inclination and sky-projected obliquity to $i_* = 139\overset{\circ}{.}2 \pm 0\overset{\circ}{.}3$ and $\lambda_* = 31\overset{\circ}{.}1 \pm 0\overset{\circ}{.}4$. These results confirm that TOI-3884 is viewed nearly pole-on, naturally explaining the weak rotational modulation observed in long-term photometry despite the presence of a large spot.
4. We find the occulted spot is extremely large and located near the stellar rotation pole. Joint modeling of all visits indicates a spot radius of $R_{\rm spot} = 0.576\,^{+0.006}_{-0.005}\,R_*$—over 3× larger than the planet's radius—and a latitude of $\phi_{\rm spot} = -84\overset{\circ}{.}69 \pm 0\overset{\circ}{.}12$, with a projected spot-coverage fraction ranging from ∼20% to 26% over the stellar rotation. Although individual visits allow some variation in inferred spot size, the spot contrast is broadly consistent across epochs, indicating that the spectral properties of the occulted region are robust to uncertainties in its detailed geometry.
5. The spot is significantly cooler than the surrounding photosphere. Comparison of the empirical contrast spectrum with 1D stellar atmosphere models indicates a spot temperature contrast of $\Delta T = 183 \pm 1$ K relative to the photosphere, corresponding to a spot temperature of $T_{\rm spot} = 2993 \pm 10$ K for a photospheric temperature of $T_{\rm phot} = 3176 \pm 10$ K.
6. The empirical spot spectrum reveals model limitations and underscores the need for direct constraints. While 1D stellar atmosphere models reproduce the observed spot contrasts reasonably well at wavelengths $\gtrsim 1\,\mu$m, we measure significantly larger contrasts at shorter wavelengths, indicating limitations in current models and/or additional structural complexity within the active region. These discrepancies demonstrate that model-dependent contamination corrections may be incomplete, particularly in the optical. The panchromatic empirical spectrum presented here therefore provides a critical benchmark for refining stellar atmosphere models and for robustly correcting transmission spectra of planets orbiting active late-type stars.

TOI-3884 b demonstrates the power of using transiting planets as spatial probes of stellar photospheres. Extending this approach to additional JWST observations of planets transiting starspots could enable the construction of an empirical library of stellar surface spectra. Such measurements will be essential for improving stellar atmosphere models, refining contamination corrections, and ultimately enabling accurate atmospheric characterization of small planets transiting active, low-mass stars.

## Acknowledgments

We thank the reviewer for the feedback that helped improve this manuscript. We thank Sabina Sagynbayeva for insightful conversations on parallel tempering and `ptemcee`. We thank Mayuko Mori for sharing spot contrasts derived from their work in M. Mori et al. (2025).

We give thanks to the program coordinators and instrument scientists at the Space Telescope, especially Michael Leveille, Glen Wahlgreen, and Tyler Baines, for assisting with observation planning and scheduling for JWST-GO-5863. Based on observations with the NASA/ESA/CSA JWST obtained from the Mikulski Archive for Space Telescopes (MAST) at the Space Telescope Science Institute, which is operated by the Association of Universities for Research in Astronomy, Incorporated, under NASA contract NAS5-03127. Support for program Nos. JWST-GO-5863 and JWST-GO-5799 was provided through grants from the STScI under NASA contract NAS5-03127.

This material is based upon work supported by the National Aeronautics and Space Administration under agreement No. 80NSSC21K0593 for the program "Alien Earths." The results reported herein benefited from collaborations and/or information exchange within NASA's Nexus for Exoplanet System Science (NExSS) research coordination network sponsored by NASA's Science Mission Directorate. This material is based upon work supported by the European Research Council (ERC) Synergy Grant under the European Union's Horizon 2020 research and innovation program (grant No. 101118581—project REVEAL).

Funding for K.B. was provided by the European Union (ERC AdG SUBSTELLAR, GA 101054354).

This work utilized the Blanca condo computing resource at the University of Colorado Boulder. Blanca is jointly funded by computing users and the University of Colorado Boulder.

*Facility:* JWST (NIRISS, NIRSpec).

*Software:* `astropy` (Astropy Collaboration et al. 2013, 2018, 2022), `NumPy` (C. R. Harris et al. 2020), `matplotlib` (J. D. Hunter 2007), `emcee` (D. Foreman-Mackey et al. 2013), `ptemcee` (D. Foreman-Mackey et al. 2013; W. D. Vousden et al. 2016), fleck (B. Morris 2020), `speclib` (B. V. Rackham 2023; B. V. Rackham & J. de Wit 2024), `chromatic` (Z. Berta-Thompson et al. 2025), `spotter` (L. Garcia et al. 2025), `ExoTiC-LD` (D. Grant & H. R. Wakeford 2024).

## Appendix A
## Detailed Data Reduction Procedures

We reduced all seven JWST visits independently with two independent pipelines: `Eureka!` (v1.2; T. J. Bell et al. 2022)[28] and `ExoTEDRF` (formerly `supreme-SPOON`, M. Radica et al. 2023).[29] In the main analysis, we adopt the ExoTEDF reductions. Here, we provide the detailed configuration choices and modifications applied in each pipeline.

### A.1. Eureka!

`Eureka!` is an end-to-end pipeline for the reduction and analysis of JWST and HST observations (T. J. Bell et al. 2022). Its workflow is structured in six stages. The first three, which we relied upon, are dedicated to calibration, correction procedures, and optimal extraction of spectroscopic data. We analyzed the data with Eureka! version 1.2, which introduced support for NIRISS data reduction. Leveraging this newly added compatibility, we independently reduced both the NIRSpec and NIRISS observations. As is standard practice for NIRSpec/G395H observations, the NRS1 and NRS2 data were reduced separately. Below, we describe the steps common to all reductions and the changes made to the default Eureka! inputs to improve the reduction quality. We used both visual inspection of the inter-stage diagnostic outputs and the median absolute deviation value for the Stage 3 spectroscopic light curve to guide our selection of optimal reduction parameters.

Stage 1 and 2 are wrappers for the `jwst` v1.18 pipeline (H. Bushouse et al. 2024). Stage 1 performs core detector-level processing, including ramp fitting, saturation flagging, linearity correction, dark current subtraction, and cosmic ray detection. Eureka! includes some targeted modifications to this stage, designed for time series observations. These include group-level background subtraction to mitigate $1/f$ noise, a custom bias correction scheme for NIRSpec/G395H, and more aggressive saturation flagging. For all nine reductions, we explored a variety of Stage 1 configurations. The parameters we modified from the default were: (1) the jump rejection threshold, raised from $4\sigma$ to $6\sigma$ as commonly done for time series observations, (2) setting the bias correction method to mean for the NIRSpec/G395H reductions, (3) enabling group-level background subtraction to remove $1/f$ noise and adjusting the relevant background region row boundaries to [6, 160] and [3, 28] for NIRISS/SOSS and NIRSpec/G395M reductions, respectively, and (4) enabling aggressive saturation flagging for the NIRSpec/G395H observations. For Stage 2, we used the default Eureka! settings, performing the standard steps of world coordinate system assignment and flat-field subtraction.

Prior to Stage 1, we manually corrected the NIRISS/SOSS observations for the well-known zodiacal background contamination. Rather than relying on Eureka!'s built-in background subtraction routines, we scaled the STScI-provided NIRISS/SOSS background model to our data before subtracting it from each individual frame. The zodiacal background model was scaled differentially across detector rows but uniformly across the step. A global scaling factor was first derived from the median count level of a background-only region of each integration, then, for each detector row, the corresponding background model row was scaled by the global scaling factor and normalized by the row's median count before being subtracted. The goal of this procedure was to preserve row-dependent structure while avoiding residual zodiacal background, which we found present when not performing the background correction with row-by-row adjustments.

Stage 3 is when Eureka! diverges markedly from the standard `jwst` pipeline. It produces a time series of stellar spectra by locating the spectral trace, correcting for its curvature, performing column-by-column background subtraction via a polynomial fit with iterative sigma clipping, and finally applying the optimal extraction algorithm of K. Horne (1986). As with Stage 1, we explore a range of configurations. Our final reductions differed from the defaults in four ways. First, on a case-by-case basis, we trimmed the subarray region of interest to retain as much of the spectral trace as possible while discarding detector regions with no signal. Second, we applied the full-frame outlier rejection routine along the time axis, while verifying that the fraction of flagged pixels remains well below 1%. Third, we increased the outlier region threshold from $4\sigma$ to $6\sigma$ in the aforementioned routine. Lastly, we adjusted the background and spectral extraction aperture half-widths. For all NIRISS reductions, we adopted half-widths of 27 and 24 pixels for the background and extraction apertures, respectively. For NIRSpec, we used apertures of 6 and 7 pixels for G395M and 8 and 9 pixels for G395H.

### A.2. ExoTEDRF

We perform a second, independent reduction using `ExoTEDRF` (A. D. Feinstein et al. 2023; M. Radica et al. 2023; M. Radica 2024). Similarly to `Eureka!`, `ExoTEDRF` provides an end-to-end reduction and analysis pipeline for JWST exoplanet time series observations. `ExoTEDRF` was initially optimized for NIRISS/SOSS; however, it now supports reduction techniques for NIRSpec/BOTS as well. Thus, we used `ExoTEDRF` to independently reduce all NIRISS/SOSS and NIRSpec/G395M and G395H observations. We describe our NIRISS/SOSS reduction, followed by NIRSpec.

#### A.2.1. NIRISS/SOSS

We performed Stage 1 corrections on our NIRISS/SOSS observations using the `jwst` pipeline with additional modifications. Stage 1 corrections that used the `jwst` pipeline include initializing the data quality flags and saturated pixel flags, superbias subtraction, reference pixel correction, dark current correction, and background correction. We correct the background by scaling the NIRISS/SOSS background model provided by STScI to our observations. This scaling is determined by scaling the model background to the counts level of a median stack of the exposures. This scaled model background is then subtracted from each integration. There are very few regions of the exposure that are not illuminated by the SOSS point-spread function. For these observations, we use the following region: $x = [230, 250]; y = [350, 550]$, as is standard within `ExoTEDRF`. After this stage, we perform $1/f$ noise correction at the group level, as group-level correction can lead to higher precision light curves with lower noise levels (A. Carter et al. 2025). We use the "scale-achromatic" $1/f$ noise correction within `ExoTEDRF`, which constructs difference images of each exposure by subtracting a median image, then subtracts the median value from each column. The scaled background is then re-added to the exposures, as it needs to be flat-field corrected. The remaining Stage 1 steps are standard with the `jwst` pipeline.

[28] https://eurekadocs.readthedocs.io/en/latest/
[29] https://exotedrf.readthedocs.io/en/latest/

For our Stage 2 processing, we again apply the standard `jwst` pipeline routines of assigning the world coordinate system and flat field subtraction. Here, we subtract the scaled STScI background model, which was described in the Stage 1 processing. We then correct for bad pixels and run the principal component analysis reconstruction step (M. Radica et al. 2025). This stage allows us to remove similar components between exposures, such as components that correspond to detector systematics. We remove Proportional Counter Array components 2, 3, and 4 based on visual inspection, as these correspond to detector-level systematics and are not astrophysical in nature. We extract our stellar spectra using a standard box extraction with an aperture width of 30 pixels.

#### *A.2.2. NIRSpec/BOTS*

Our NIRSpec/G395M and G395H reduction is nearly identical to our NIRISS/SOSS reduction, with some modifications. First, we do not need to correct for the same structured background that is seen in NIRISS/SOSS. Second, in our Stage 2 processing, we perform wavelength correction in the instance that the target is not perfectly centered in the slit. Third, we perform additional $1/f$ noise correction at the integration level. This removes any potential residual noise that may still be present, even after the aforementioned reduction steps. Again, we extract our stellar spectra using a standard box extraction with an aperture width of 8 pixels. This was performed independently for the NIRSpec/G395H NRS1 and NRS2 detectors.

## Appendix B
## Flare Outburst Events

Here, we illustrate the large flare outburst events observed during the early JWST visits. Figure 6 compares the NIRISS order 2 white light curves to light curves extracted around H$\alpha$, highlighting the chromospheric origin and temporal structure of these events.

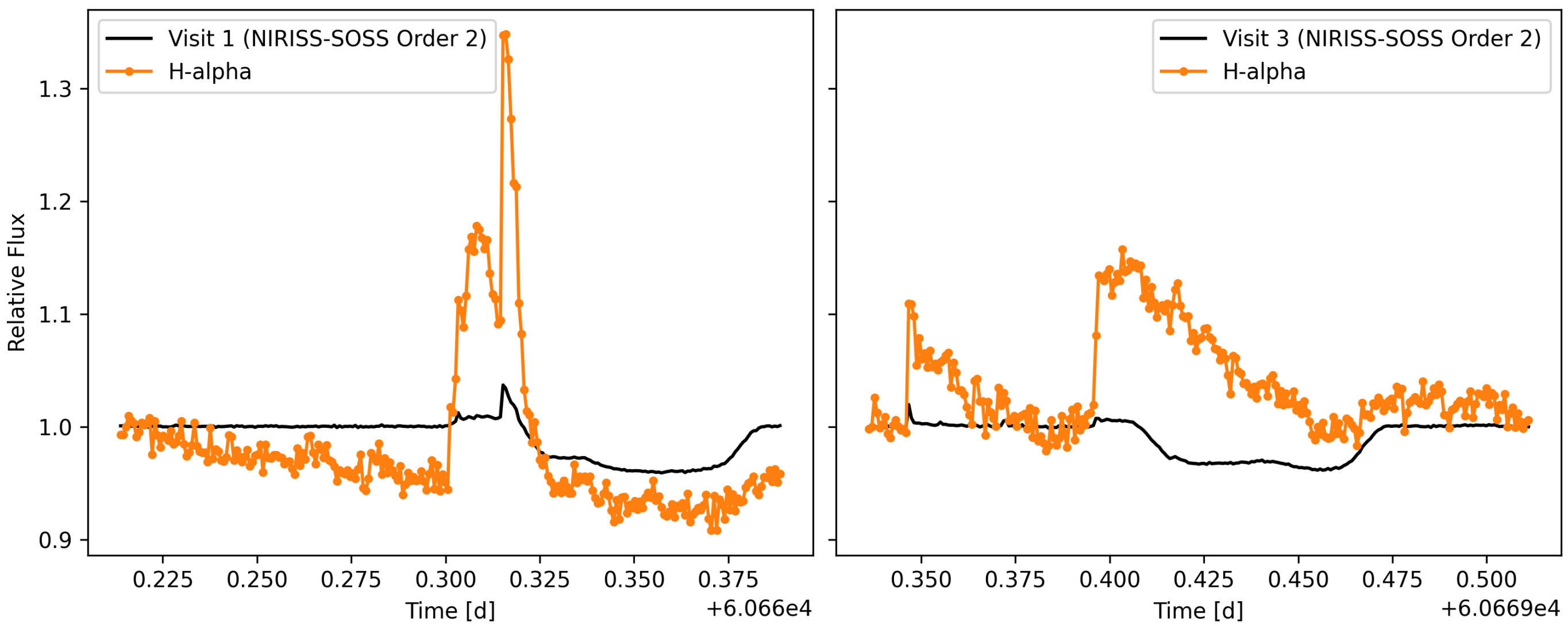

**Figure 6.** White light curves from the JWST-GO-5863 observations of TOI-3884 b, taken with NIRISS/SOSS (order 2, 0.63–1.10 $\mu$m), shown in black. Light curves extracted around H$\alpha$ emission (0.655–0.6565 $\mu$m) are overplotted in orange.

# Appendix C
# Details of the Individual Light-curve Analysis

To verify our results from Section 3.2, we performed a similar light-curve analysis on each JWST visit individually. In this appendix, we describe the details of the white light-curve and spectroscopic light-curve fits. In this analysis, we opted to include Visits 1 and 3, but mask the obvious flaring regions (Figure 2), leaving partial transits and spot crossings to fit. When we fit Visit 3 (rather than removing as in Section 3.2), we find discrepant $R_{\mathrm{spot}}$ and $C$ values (Figure 4), further justifying our decision to remove the visit from our joint analysis.

## C.1. White Light-curve Fits

For each of the seven white light curves (order 1 for NIRISS and NRS1 for NIRSpec/G395H), we fixed the wavelength-independent stellar and planetary parameters to the results from the NIRSpec joint fit (Table 2), while allowing the spot to vary. Therefore, we fit only for the spot parameters ($R_{\mathrm{spot}}$, $\phi_{\mathrm{spot}}$, and $\lambda_{\mathrm{spot}}$) and wavelength-dependent quantities: $R_{\mathrm{p}}/R_{*}$, $q_1$, $q_2$, $C$, $p$, and $\ln \sigma_{\mathrm{infl}}$. The parameter posteriors were sampled with ptemcee as in Section 3.1.5, but with 25 temperatures, 30 walkers, and a burn-in of 5000 steps followed by 10,000 steps. In all cases, chain lengths exceeded the integrated autocorrelation time, indicating convergence.

## C.2. Spectroscopic Light Curves

Similarly to Section 3.2.3, we fit each of the $R = 100$ light curves, adopting the white-light best-fit values for each visit from Section C.1. We opted not to include Visits 1 and 3 due to a lack of constraint on spot radius and contrast (Figure 4). As in Section 3.2.3, we fitted only for the wavelength-dependent variables and flare amplitudes in Visit 5. The $R = 100$ spot contrast spectra for each visit are displayed in Figure 7.

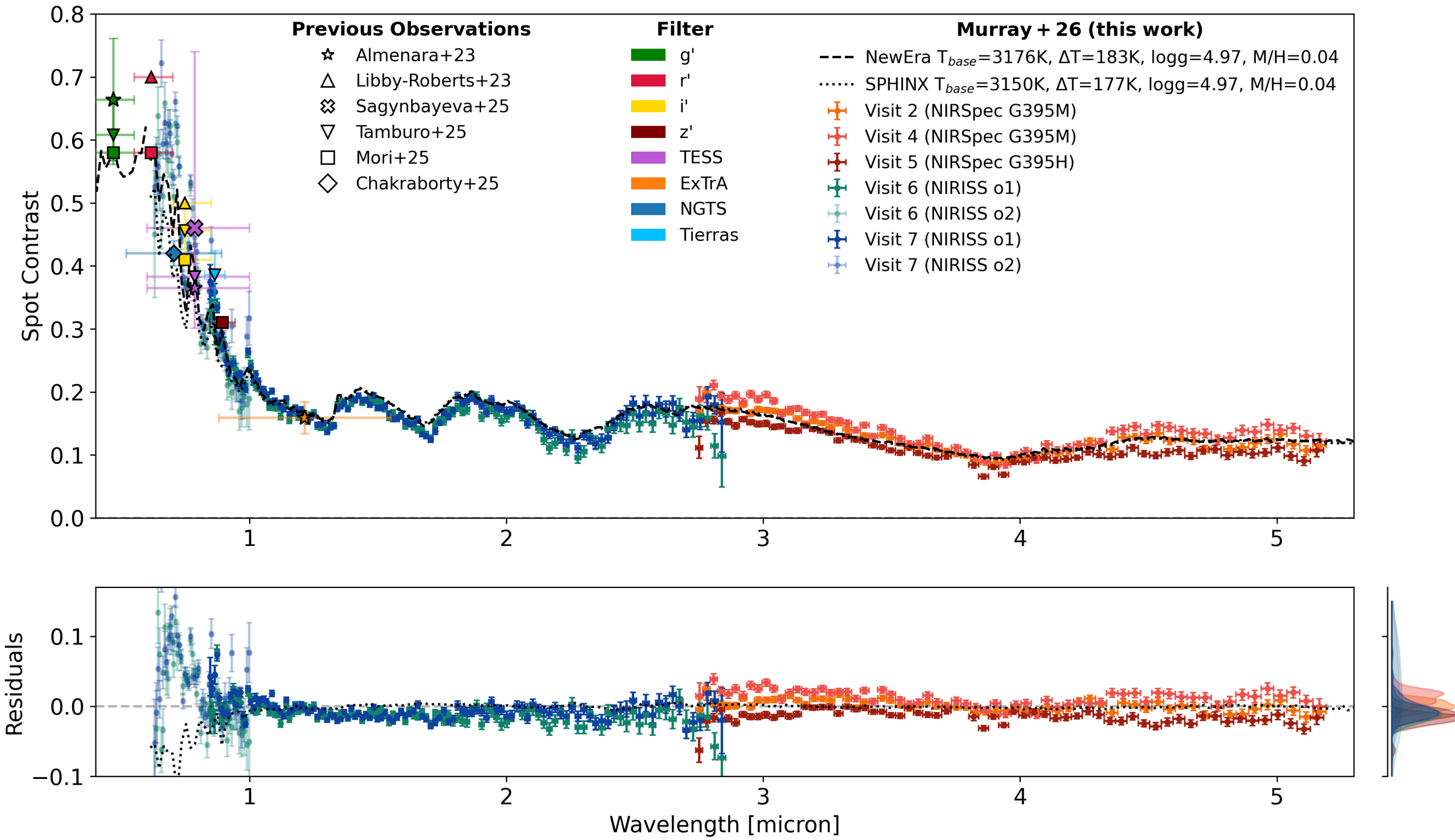

**Figure 7.** Similar to Figure 5, but for the individually fit light curves described in Section 3.3 and Appendix C. The best-fit NewEra/SPHINX models for the self-consistent joint model are overplotted in black.

## ORCID iDs

C. A. Murray https://orcid.org/0000-0001-8504-5862
L. Garcia https://orcid.org/0000-0002-4296-2246
B. V. Rackham https://orcid.org/0000-0002-3627-1676
Z. Berta-Thompson https://orcid.org/0000-0002-3321-4924
A. D. Feinstein https://orcid.org/0000-0002-9464-8101
S. J. Mercier https://orcid.org/0000-0002-0962-7585
B. Charnay https://orcid.org/0000-0003-0977-6545
L. Hebb https://orcid.org/0000-0003-1263-8637
J. E. Libby-Roberts https://orcid.org/0000-0002-2990-7613
Y. Rotman https://orcid.org/0000-0003-4459-9054
A. Stephens https://orcid.org/0009-0009-7798-8700
M. Timmermans https://orcid.org/0009-0008-2214-5039
L. Welbanks https://orcid.org/0000-0003-0156-4564
K. Barkaoui https://orcid.org/0000-0003-1464-9276
Caleb I. Cañas https://orcid.org/0000-0003-4835-0619
M. Delamer https://orcid.org/0000-0003-1439-2781
E. Ducrot https://orcid.org/0000-0002-7008-6888
S. Kanodia https://orcid.org/0000-0001-8401-4300
S. Mahadevan https://orcid.org/0000-0001-9596-7983
J. P. Ninan https://orcid.org/0000-0001-8720-5612
J. de Wit https://orcid.org/0000-0003-2415-2191